\documentclass[apj, 8pt]{emulateapj-rtx4}
\usepackage{color}
\usepackage{amsmath}
\usepackage[dvipdfm]{hyperref}
\hypersetup{
	bookmarks=true,
	bookmarksnumbered=true,
	bookmarksopen=true,
	colorlinks   = true,
	citecolor    = blue,
	pdftitle = {Evolution of Galaxy Size},
	pdfauthor = {Takatoshi Shibuya},
	pdfkeywords={cosmology: observations --- early universe --- galaxies: formation --- galaxies: high-redshift}
	pdfpagemode=UseThumbs
}

\renewcommand\thefootnote{\arabic{footnote}}

\slugcomment{Submitted to ApJS March 25, 2015; accepted May 6, 2015}

\shorttitle{Evolution of Galaxy Size at $z=0-10$}
\shortauthors{T. Shibuya et al.}

\begin{document}

\title{Morphologies of $\sim 190,000$ Galaxies at $\lowercase{z}=0-10$ 
Revealed with HST Legacy Data \\I. Size Evolution}

%% AUTHOR
%%%%%%%%%%%%%%%%%%%%%%%%%%%%%%%%%%%%%%%%%%%%%%%%

%\author{Takatoshi SHIBUYA\altaffilmark{1,2} et al.}
\author{Takatoshi Shibuya\altaffilmark{1,2}, Masami Ouchi\altaffilmark{1,3}, and Yuichi Harikane\altaffilmark{1,4}}
\email{shibyatk\_at\_icrr.u-tokyo.ac.jp}

\altaffiltext{1}{Institute for Cosmic Ray Research, The University of Tokyo, 5-1-5 Kashiwanoha, Kashiwa, Chiba 277-8582, Japan}
\altaffiltext{2}{Center for Computational Sciences, The University of Tsukuba, 1-1-1 Tennodai, Tsukuba, Ibaraki 305-8577 Japan}
\altaffiltext{3}{Kavli Institute for the Physics and Mathematics of the Universe (Kavli IPMU, WPI), University of Tokyo, Kashiwa, Chiba 277-8583, Japan}
\altaffiltext{4}{Department of Physics, Graduate School of Science, The University of Tokyo, 7-3-1 Hongo, Bunkyo, Tokyo 113-0033, Japan}

%% ABSTRACT
%%%%%%%%%%%%%%%%%%%%%%%%%%%%%%%%%%%%%%%%%%%%%%%%

\begin{abstract}

We present redshift evolution of galaxy effective radius $r_e$ obtained from the $\!${\it Hubble Space Telescope} ($\!${\it HST}) samples of $\sim 190,000$ galaxies at $z=0-10$. Our $\!${\it HST} samples consist of $176,152$ photo-$z$ galaxies at $z=0-6$ from the 3D-HST+CANDELS catalogue and $10,454$ Lyman break galaxies (LBGs) at $z=4-10$ identified in CANDELS, HUDF09/12, and HFF parallel fields, providing the largest data set to date for galaxy size evolution studies. We derive $r_e$ with the {\it same} technique over the wide-redshift range of $z=0-10$, evaluating the optical-to-UV morphological {\it K}-correction and the selection bias of photo-$z$ galaxies+LBGs as well as the cosmological surface brightness dimming effect. We find that $r_e$ values at a given luminosity significantly decrease towards high-$z$, regardless of statistics choices (e.g. $r_e\propto(1+z)^{-1.10\pm0.06}$ for median). For star-forming galaxies, there is no evolution of the power-law slope of the size-luminosity relation and the median S\'ersic index ($n\sim 1.5$). Moreover, the $r_e$-distribution is well represented by log-normal functions whose standard deviation $\sigma_{\ln{r_e}}$ does not show significant evolution within the range of $\sigma_{\ln{r_e}} \sim0.45-0.75$. We calculate the stellar-to-halo size ratio from our $r_e$ measurements and the dark-matter halo masses estimated from the abundance matching study, and obtain a nearly constant value of $r_e/r_{\rm vir}=1.0-3.5$\% at $z=0-8$.  The combination of the $r_e$-distribution shape+standard deviation, the constant $r_e/r_{\rm vir}$, and $n\sim1.5$ suggests a picture that typical high-$z$ star-forming galaxies have disk-like stellar components in a sense of dynamics and morphology over cosmic time of $z\sim 0-6$. If high-$z$ star-forming galaxies are truly dominated by disks, the $r_e/r_{\rm vir}$ value and the disk formation model indicate that the specific angular momentum of the disk normalized by the host halo is $j_{\rm d}/m_{\rm d}\simeq 0.5-1$. These are statistical results for galaxies' major stellar components, and the detailed study of clumpy sub-components is presented in the paper II.
 
\end{abstract}

% clearly decreases from $z\sim0$ at least up to $\sim6$ regardless of whether we use average, median, or modal $r_e$. The median $r_e$ evolves with $\propto(1+z)^{-1.12\pm0.05}$. 
%We additionally confirm a trend of smaller $r_e$ with a bluer UV slope $\beta$, a constant ratio of $r_e$ to virial radius of DM halos, $r_e/r_{\rm vir}\sim1-2$\%, and a gradual increase of SFR surface density from $z\sim0$ to $\sim4$ with no evolution at $z\gtrsim4$, all of which do not strongly depend on UV luminosity. 
%This indicates that the $r_e$ standard deviation may result from intrinsic unobservable structures in high-$z$ proto-disks such as star-forming clumps and/or a variation of spin parameter for host DM halos. 


%% INTRODUCTION: 
%%%%%%%%%%%%%%%%%%%%%%%%%%%%%%%%%%%%%%%%%%%%%%%%

\section{INTRODUCTION}

Galaxy sizes offer a variety of invaluable insights into the galaxy formation and evolution. A slope of size-stellar mass (or luminosity) relation, a size growth rate, and a size distribution are key quantities for understanding developments of galaxy morphology and properties of host dark matter (DM) halos.
%For example, a typical size of galaxy disks is expected to evolve as a function of $H(z)^{-2/3}$ and $H(z)^{-1}$ in the case of a fixed virial mass or a fixed circular velocity of DM halos, respectively, with redshift, where $H(z)$ is the Hubble parameter, according to analytical models \cite[e.g., ][]{2004ApJ...600L.107F}.
%

Studies of high-$z$ galaxy sizes show substantial progresses by observations of the $\!${\it Hubble Space Telescope} ($\!${\it HST}) which is capable of high spatial resolution imaging. Galaxy sizes defined by the effective radius, $r_e$, have extensively been measured with Advanced Camera for Survey (ACS) and Wide Field Camera 3/IR-channel onboard $\!${\it HST} for massive galaxies at $0\lesssim z\lesssim3$ \citep[e.g., ][]{2014ApJ...788...28V} and $z\gtrsim3-4$ Lyman break galaxies (LBGs) selected in the dropout technique \citep{1999ApJ...519....1S} \citep[e.g., ][]{2006ApJ...650...18T,2007ApJ...671..285T,2009ApJ...705..255T,2007ApJ...654..172D,2012AA...547A..51G,2013ApJ...765...68H,2013MNRAS.428.1088M}. However, these studies, particularly at high-$z$, do not reach agreement on the size growth rate. \citet{2010ApJ...709L..21O} have reported that the average size evolves according to roughly $r_e\propto(1+z)^{-1}$ based on a $z\sim7$ LBG sample in an early-epoch data of their $\!${\it HST} survey \citep[see also e.g., ][]{2004ApJ...611L...1B,2014arXiv1406.1180H}. On the other hand, \citet{2008ApJ...673..686H} have argued that the average size scales as $r_e\propto(1+z)^{-1.5}$ using LBGs at $z\sim2-6$ \citep[see also e.g., ][]{2004ApJ...600L.107F}. Some studies have provided results of a growth rate falling between these two growth rates \citep[e.g., ][]{2012ApJ...756L..12M,2013ApJ...777..117M,2013ApJ...777..155O}. Moreover, \citet{2014arXiv1409.1832C} have suggested no significant evolution of $\!${\it typical} galaxy sizes, if one does not use average but modal values of size distribution for representative radii at a given redshift. These discrepancies in the evolutionary trend would be attributed to small $\!${\it HST} samples at $z\gtrsim3-4$ and/or potential biases caused by heterogenous samples and measurements taken from the literature.

\begin{deluxetable*}{ccccccc}
\setlength{\tabcolsep}{0.35cm} 
\tabletypesize{\scriptsize}
\tablecaption{Limiting Magnitudes of HST Images for Size Analyses}
\tablehead{ & \multicolumn{6}{c}{$15\sigma$ ($5\sigma$) Depth}\\
\colhead{Field} & \colhead{$V_{606}$} & \colhead{$I_{814}/z_{850}$} & \colhead{$J_{125}$} & \colhead{$H_{160}$} & \colhead{$Y_{098}Y_{105}J_{125}H_{160}$\tablenotemark{a}} & \colhead{$J_{125}H_{160}$\tablenotemark{b}}\\
\colhead{(1)}& \colhead{(2)}& \colhead{(3)}& \colhead{(4)} &  \colhead{(5)}& \colhead{(6)}& \colhead{(7)} } 

\startdata

HUDF09+12 & 29.3 (30.5) & 28.1 (29.3) & 28.7 (29.9) & 28.7 (29.9) & 29.4 (30.6) & 29.2 (30.4) \\ 
HUDF09-P1 & 29.4 (30.6) & 28.7 (29.9) & 28.0 (29.2) & 27.8 (29.0) & 28.4 (29.6) & 28.2 (29.4) \\ 
HUDF09-P2 & 28.2 (29.4) & 27.5 (28.7) & 28.2 (29.4) & 28.0 (29.2) & 28.6 (29.8) & 28.5 (29.7) \\
GOODS-S Deep & 27.6 (28.8) & 27.6 (28.8) & 27.2 (28.4) & 27.1 (28.3) & 27.8 (29.0) & 27.5 (28.7) \\
GOODS-S Wide & 27.6 (28.8) & 27.2 (28.4) & 26.6 (27.8) & 26.4 (27.6) & 27.1 (28.3) & 26.9 (28.1) \\
GOODS-N Deep & 27.6 (28.8) & 29.3 (30.5) & 27.1 (28.3) & 26.9 (28.1) & 27.3 (28.5) & 27.4 (28.6) \\
GOODS-N Wide & 27.5 (28.7) & 28.7 (29.9) & 26.4 (27.6) & 26.3 (27.5) & 26.9 (28.1) & 26.7 (27.9) \\
UDS & 27.0 (28.2) & 27.0 (28.2) & 26.3 (27.5) & 26.4 (27.6) & 26.7 (27.9) & \nodata \\
COSMOS & 27.1 (28.3) & 26.8 (28.0) & 26.4 (27.6) & 26.4 (27.6) & 26.7 (27.9) & \nodata \\
AEGIS & 27.1 (28.3) & 26.6 (27.8) & 26.4 (27.6) & 26.5 (27.7) & 26.8 (28.0) & \nodata \\
HFF-Abell2744P & 27.9 (29.1) & 27.6 (28.8) & 27.6 (28.8) & 27.6 (28.8) & 28.1 (29.3) & 27.9 (29.1) \\
HFF-MACS0416P & 27.7 (28.9) & 27.6 (28.8) & 27.9 (29.1) & 27.8 (29.0) & 28.3 (29.5) & 28.1 (29.3) \\ \hline
PSF FWHM\tablenotemark{c} & $0.\!\!^{\prime\prime}08$ & $0.\!\!^{\prime\prime}09$ & $0.\!\!^{\prime\prime}12$ & $0.\!\!^{\prime\prime}18$ & $0.\!\!^{\prime\prime}18$ & $0.\!\!^{\prime\prime}18$

\enddata

\tablecomments{Columns: (1) Field. (2)-(7) Limiting magnitudes defined by a $15\sigma$ ($5\sigma$ in parentheses) sky noise in a $0.\!\!^{\prime\prime}35$-diameter aperture. }
\tablenotetext{a}{Stacked image of $Y_{098}Y_{105}J_{125}H_{160}$-bands for LBGs at $z\sim4, 5, 6$, and $7$. The $JH_{140}$ image is also included for LBGs at $z\sim7-8$ in the HUDF09+12 field. }
\tablenotetext{b}{Stacked image of $J_{125}$- and $H_{160}$-bands for LBGs at $z\sim8$. }
\tablenotetext{c}{Typical size of PSF FWHMs. }
\label{table_hst_images}
\end{deluxetable*}

The two size growth rates of $r_e\propto(1+z)^{-1.5}$ and $r_e\propto(1+z)^{-1}$ correspond to the cases of a fixed virial mass and circular velocity of DM halos, respectively, if the stellar-to-halo size ratio (SHSR) is constant over the redshift range. Assuming the constant SHSR, a number of studies discuss the evolution of host DM halos with the size growth rates \cite[e.g., ][]{2004ApJ...600L.107F,2008AJ....135..156H}. However, the evolution of SHSR is not well understood. Recently, SHSRs have been estimated observationally with results of abundance matching techniques \citep[e.g., ][]{2010ApJ...717..379B,2013ApJ...770...57B} for galaxies at $z\sim0$ \citep{2013ApJ...764L..31K} and at $z\sim2-10$ \citep{2014arXiv1410.1535K}. \citet{2014arXiv1410.1535K} conclude that virtually constant value of SHSR, $3.3\pm 0.1$\%, over the wide redshift range. Galaxy disk formation models of e.g., \citet{1983IAUS..100..391F,2002ASPC..273..289F,1987ApJ...319..575B,1998MNRAS.295..319M} predict that galaxy disks acquire an angular momentum from its host DM halo trough tidal torques during the formation of these systems, leading to the proportionality between the two sizes. The SHSR values provide us with information about the DM spin parameter and the fraction of specific angular momentum transferred from DM halos to the central galaxy disks \citep[e.g., ][]{1998MNRAS.295..319M}.

\begin{deluxetable*}{ccccccc}
\setlength{\tabcolsep}{0.35cm} 
\tabletypesize{\scriptsize}
\tablecaption{Number of Photo-$z$ Galaxies for our Size Measurements}
\tablehead{ & \multicolumn{6}{c}{$N_{\tt GALFIT}/N_{\rm SFG, QGs}$}\\
\colhead{Field} & \colhead{$z=0-1$} & \colhead{$z=1-2$\tablenotemark{b}} & \colhead{$z=2-3$\tablenotemark{b}} & \colhead{$z=3-4$} & \colhead{$z=4-5$} & \colhead{$z=5-6$}\\
\colhead{(1)}& \colhead{(2)}& \colhead{(3)}& \colhead{(4)} &  \colhead{(5)}& \colhead{(6)}& \colhead{(7)} } 

\startdata
\multicolumn{7}{c}{SFGs, $r_e^{\rm Opt}(4500-8000$\AA)} \\\hline
HUDF09+12 & 294 (397) & 368 (611) & 8 (157) & \nodata & \nodata & \nodata  \\
HUDF09-P\tablenotemark{a} & 2168 (3467) & 2024 (4707) & 66 (1451) & \nodata & \nodata & \nodata  \\
GOODS-S Deep & 1753 (2793) & 1402 (3847) & 37 (1296) & \nodata & \nodata & \nodata  \\
GOODS-S Wide & 2790 (4724) & 2267 (6313) & 66 (2518) & \nodata & \nodata & \nodata  \\
GOODS-N Deep & 3270 (5106) & 1778 (5168) & 71 (2094) & \nodata & \nodata & \nodata  \\
GOODS-N Wide & 3903 (5939) & 2731 (6611) & 139 (2477) & \nodata & \nodata & \nodata  \\
UDS & 5157 (9175) & 5433 (15771) & 209 (6094) & \nodata & \nodata & \nodata  \\
AEGIS & 6441 (11074) & 5833 (13943) & 278 (6418) & \nodata & \nodata & \nodata  \\
COSMOS & 6856 (11385) & 3594 (9754)  & 179 (3915)  & \nodata  & \nodata & \nodata  \\ \hline
$N_{\rm total}(z)$ & 32632 (54060) & 25430 (66725) & 1053 (26420) & \nodata & \nodata & \nodata \\
$N_{\rm total}$ & \multicolumn{6}{c}{59115 (147205)} \\\hline
 & \multicolumn{6}{c}{} \\
\multicolumn{7}{c}{SFGs, $r_e^{\rm UV}(1500-3000$\AA)} \\\hline
HUDF09+12 & \nodata & 145 (611) & 79 (157) & 34 (69) & 19 (33) & 12 (26) \\
HUDF09-P\tablenotemark{a} & \nodata & 777 (4707) & 624 (1451) & 432 (936) & 160 (453) & 102 (177) \\
GOODS-S Deep & \nodata & 776 (3847) & 633 (1296) & 348 (696) & 101 (347) & 40 (702) \\
GOODS-S Wide & \nodata & 1297 (6313) & 1154 (2518) & 535 (1147) & 138 (487) & 44 (213) \\
GOODS-N Deep & \nodata & 1235 (5168) & 784 (2094) & 389 (987) & 154 (446) & 66 (174) \\
GOODS-N Wide & \nodata & 1711 (6611) & 1114 (2477) & 412 (962) & 165 (516) & 47 (167) \\
UDS & \nodata & 2730 (15771) & 1747 (6094) & 678 (2266) & 180 (716) & 52 (176) \\
AEGIS & \nodata & 3158 (13943) & 2182 (6418) & 952 (2768) & 228 (873) & 84 (281) \\
COSMOS & \nodata & 2413 (9754)  & 1642 (3915) & 939 (2048) & 192 (765) & 54 (296) \\ \hline
$N_{\rm total}(z)$ & \nodata & 14242 (66725) & 9959 (26420) & 4719 (11879) & 1337 (4636) & 501 (1796) \\
$N_{\rm total}$ & \multicolumn{6}{c}{30765 (165517)} \\ \hline\hline
 & \multicolumn{6}{c}{} \\
\multicolumn{7}{c}{QGs, $r_e^{\rm Opt}(4500-8000$\AA)\tablenotemark{c}} \\\hline
HUDF09+12 & 323 (743) & 133 (458) & 2 (98) & \nodata & \nodata & \nodata  \\
HUDF09-P\tablenotemark{a} & 267 (637) & 113 (365) & 1 (72) & \nodata & \nodata & \nodata  \\
GOODS-S Deep & 193 (447) & 85 (261) & 1 (46) & \nodata & \nodata & \nodata  \\
GOODS-S Wide & 369 (895) & 115 (444) & 4 (81) & \nodata & \nodata & \nodata  \\
GOODS-N Deep & 259 (623) & 86 (346) & 2 (98) & \nodata & \nodata & \nodata  \\
GOODS-N Wide & 272 (744) & 110 (382) & 2 (147) & \nodata & \nodata & \nodata  \\
UDS & 320 (1375) & 221 (933) & 7 (207) & \nodata & \nodata & \nodata  \\
AEGIS & 387 (1369) & 270 (824) & 6 (272) & \nodata & \nodata & \nodata  \\
COSMOS & 890 (1738) & 170 (602)  & 7 (127)  & \nodata  & \nodata & \nodata  \\ \hline
$N_{\rm total}(z)$ & 3013 (7934) & 1190 (4250) & 31 (1076) & \nodata & \nodata & \nodata \\
$N_{\rm total}$ & \multicolumn{6}{c}{4234 (13260)} 
\enddata

\tablecomments{Columns: (1) Field. (2)-(7) Number of the photo-$z$ galaxies that have $S/N\geq15$ and reliable {\tt GALFIT} outputs in each redshift range. The value in parentheses is the number of the photo-$z$ galaxies in the parent sample. }
\tablenotetext{a}{Total number of objects in the HUDF09-P1 and  HUDF09-P2 fields. }
\tablenotetext{b}{The actual redshift range is $2\leq z\leq2.1$ ($1.2\leq z\leq2$) for the $r_e^{\rm Opt}$ ($r_e^{\rm UV}$) measurement. }
\tablenotetext{c}{The numbers of QGs with $r_e^{\rm UV}$ are not shown here due to the rarity at $z\gtrsim2-3$ and the UV faintness. }
\label{table_number_sfg}
\end{deluxetable*}

\begin{deluxetable*}{ccccccc}
\setlength{\tabcolsep}{0.35cm} 
\tabletypesize{\scriptsize}
\tablecaption{Number of LBGs for our Size Measurements}
\tablehead{ & \multicolumn{6}{c}{$N_{\tt GALFIT}/N_{\rm LBG}$}\\
\colhead{Field} & \colhead{$z\sim4$} & \colhead{$z\sim5$} & \colhead{$z\sim6$} & \colhead{$z\sim7$} & \colhead{$z\sim8$} & \colhead{$z\sim10$}\\
\colhead{(1)}& \colhead{(2)}& \colhead{(3)}& \colhead{(4)} &  \colhead{(5)}& \colhead{(6)}& \colhead{(7)} } 

\startdata
HUDF09+12 & 160 (348) & 43 (130) & 26 (86) & 13 (50) & 9 (24) & 0 (2) \\
HUDF09-P1 & \nodata & 41 (95) & 12 (30) & 2 (9) & 2 (7) & 0 (0) \\
HUDF09-P2 & \nodata & 30 (90) & 8 (37) & 4 (23) & 0 (16) & 0 (0) \\
GOODS-S Deep & 1046 (1872) & 292 (696) & 122 (311) & 55 (203) & 11 (57) & 1 (1) \\
GOODS-S Wide & 294 (510) & 73 (142) & 20 (51) & 9 (31) & 3 (21) & 0 (0) \\
GOODS-N Deep & 868 (1655) & 279 (630) & 48 (135) & 35 (111) & 12 (28) & 1 (2) \\
GOODS-N Wide & 522 (800) & 106 (222) & 25 (68) & 12 (231) & 3 (28)  & 1 (1) \\
UDS & \nodata & 152 (310) & 39 (65) & 12 (25) & \nodata & \nodata  \\
AEGIS & \nodata & 189 (381) & 47 (101) & 11 (28) & \nodata & \nodata  \\
COSMOS & \nodata & 209 (348)  & 40 (80)  & 11 (27) & \nodata & \nodata  \\
HFF-Abell2744P & \nodata & 15 (37) & 12 (26) & 4 (7) & 2 (7) & \nodata  \\
HFF-MACS0416P & \nodata & 30 (134) & 23 (106) & 5 (18) & 4 (10) & \nodata \\ \hline
$N_{\rm total}(z)$ & 2890 (5185) & 1459 (3215) & 422 (1096) & 173 (763) & 46 (195) & 3 (6) \\
$N_{\rm total}$ & \multicolumn{6}{c}{4993 (10454)}
\enddata

\tablecomments{Columns: (1) Field. (2)-(7) Number of the LBGs that have $S/N\geq15$ and reliable {\tt GALFIT} outputs in each redshift range. The value in parentheses is the number of LBGs in the parent sample.}
\label{table_number_lbg}
\end{deluxetable*}

Additionally, the size-stellar mass relation and the scatter of size distribution present independent evidence for the picture of galaxy disk formation \citep[e.g., ][]{1983IAUS..100..391F,2002ASPC..273..289F,2003MNRAS.343..978S,2001ApJ...555..240B}. \citet{2014ApJ...788...28V} have revealed that the slope of size-stellar mass relation and the scatter do not significantly evolve at $0\lesssim z\lesssim3$ in a systematic structural analysis for large samples of star-forming galaxies (SFGs) and quiescent galaxies (QGs) with a photometric redshift (photo-$z$). The constant values of these quantities strongly suggest that the sizes of SFGs are determined by their host DM halos. However, the controversial results of the slope and scatter evolution are obtained at $z\gtrsim3-4$ \citep[e.g., ][]{2013ApJ...765...68H,2014arXiv1409.1832C}, probably due to large statistical uncertainties given by the small galaxy samples. An analysis with a large LBG sample would reveal the galaxy structure evolution up to $z\sim10$ with no significant statistical uncertainties, and allow us to understand disk formation mechanisms, internal star formation, and morphological evolution over cosmic time.

\begin{figure}[t!]
  \begin{center}
    \includegraphics[width=83mm]{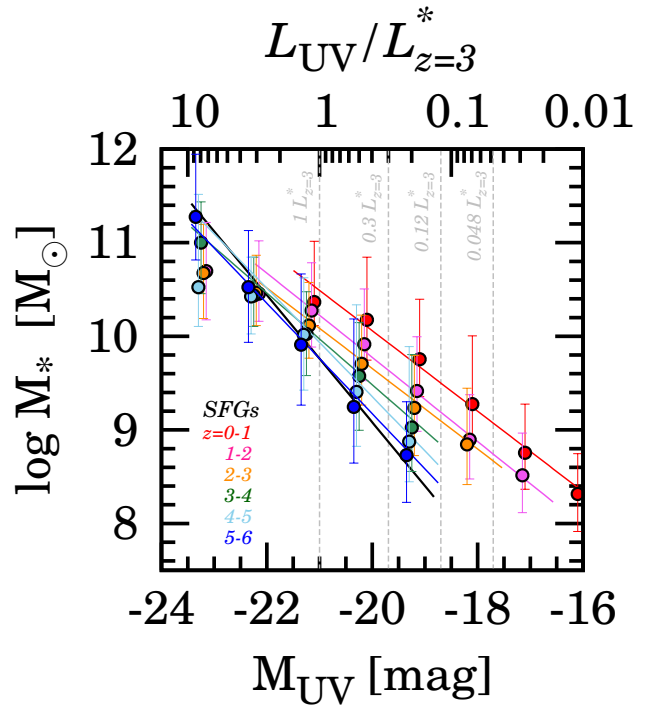}
  \end{center}
  \caption[]{{\footnotesize Relation between UV magnitude and stellar mass for the SFGs at $z\sim0-6$. The circles with error bars indicate SFGs at $z=0-1$ (red), $z=1-2$ (magenta), $2-3$ (orange), $3-4$ (green), $4-5$ (cyan), and $5-6$ (blue). The colored solid lines present the best-fit $M_*$-$M_{\rm UV}$ relation at each $z$ bin with the color coding same as the one of the circles.The black solid line denotes the $M_*$-$M_{\rm UV}$ relation for LBGs at $z\sim4$ in \citet{2011ApJ...735L..34G}. The top x-axis provides the corresponding UV luminosity in units of $L_{z=3}^*$. The error bars denote the 16th and 84th percentiles of distribution. The vertical dashed lines denote thresholds of $L_{\rm UV}$ bins, $1$, $0.3$, $0.12$, and $0.048\, L_{\rm UV}/L_{z=3}^*$, from left to right. }}
  \label{fig_muv_mass_sfg}
\end{figure}

%% EXPLANATION OF THIS PAPER
%%%%%%%%%%%%%%%%%%%%%%%%%%%%%%%%%%%%%%%%%%%%%%%%

In this paper, we systematically investigate redshift evolution of galaxy sizes with an unprecedentedly large sample of $186,603$ galaxies at $z=0-10$ made from the $\!${\it HST} deep data of extra-galactic legacy surveys. We assess effects of morphological {\it K}-correction, statistics choice, and sample selection bias with the galaxies at $z\lesssim 4$, and then extend our systematic morphological measurements to $z\gtrsim 4$. This paper has the following structure. In Section \ref{sec_data}, we describe the details of our $\!${\it HST} galaxy samples. Section \ref{sec_analysis_size} presents methods for estimating galaxy sizes. In Section \ref{sec_bias}, we evaluate the morphological {\it K} corrections, statistics-choice dependences, and selection biases. We show the redshift evolution of size-relevant physical quantities in Section \ref{sec_results}. Section \ref{sec_discussion} discusses the implications for galaxy formation and evolution with results of our structural analyses. We summarize our findings in Section \ref{sec_conclusion}.

Throughout this paper, we adopt the concordance cosmology with $(\Omega_m, \Omega_\Lambda, h)=(0.3, 0.7, 0.7)$, \citep{2011ApJS..192...18K}. All magnitudes are given in the AB system \citep{1983ApJ...266..713O}. We refer to the HST F606W, F775W, F814W, F850LP, F098M, F105W, F125W, F140W, and F160W filters as $V_{606}, i_{775}, I_{814}, z_{850}, Y_{098}, Y_{105}, J_{125}, JH_{140}$, and $H_{160}$, respectively.

%% SAMPLE
%%%%%%%%%%%%%%%%%%%%%%%%%%%%%%%%%%%%%%%%%%%%%%%%
\section{Data and Samples}\label{sec_data}

We make use of the following two galaxy samples constructed from the deep optical and near-infrared imaging data taken by $\!${\it HST} deep extra-galactic legacy surveys whose limiting magnitudes and PSF FWHM sizes are summarized in Table \ref{table_hst_images}. In the last subsection, we explain the stellar masses of the sample galaxies.

\subsection{Sample of Photo-$z$ Galaxies at $z=0-6$\\ in 3D-HST+CANDELS}\label{subsec_3dhst}

The first sample is made of $176,152$ $\!${\it HST}/WFC3-IR detected galaxies with photometric redshifts (hereafter photo-$z$ galaxies) at $z=0-6$ taken from \citet{2014arXiv1403.3689S}. These galaxies are identified in five Cosmic Assembly Near-infrared Deep Extragalactic Legacy Survey (CANDELS)  fields \citep{2011ApJS..197...35G, 2011ApJS..197...36K}, and detected in stacked images of $J_{125}, JH_{140}$, and $H_{160}$ bands of WFC3/IR, which yields roughly a stellar mass-limited sample. The photometric properties and the results of spectral energy distribution (SED) fitting for all the sources are summarized in \citet{2014arXiv1403.3689S}. The $\!${\it HST} images and catalogues are publicly released at the 3D-HST website\footnote{http://3dhst.research.yale.edu/Home.html}. The catalogues include the spectroscopic redshifts on the basis of the $\!${\it HST}/WFC3 G141 grism observation \citep{2012ApJ...758L..17B}. We use galaxies whose physical quantities and photometric redshifts are well derived from SED fitting (specifically, sources with {\tt use\_phot}$=1$ in the public catalogues). Tables \ref{table_number_sfg} summarizes the number of galaxies at each redshift in the photo-$z$ galaxy sample that we use. In this paper, we assume \citet{1955ApJ...121..161S} initial mass function (IMF). To obtain the Salpeter IMF values of stellar masses ($M_*$) and star formation rates (SFRs), we multiply the \citet{2003PASP..115..763C} IMF values from the \citet{2014arXiv1403.3689S} catalogue by a factor of $1.8$. We divide the sample of photo-$z$ galaxies at $z=0-4$ into two subsamples of star-forming galaxies (SFGs) and quiescent galaxies (QGs) by the rest-frame {\it UVJ} color criteria of \citet{2013ApJ...777...18M}. Because the {\it UVJ} color criteria are not tested for $z>~4$ sources, we do not apply these color criteria to the photo-$z$ galaxies at $z>4$. \citet{2013ApJ...777...18M} find that the QG fraction is small, 10\%, at $z\sim 3.5$, and it is likely that a QG fraction at the early cosmic epoch of $z>4$ is negligibly small, perhaps $<10$\%. We thus regard all of the $z>4$ photo-$z$ galaxies as SFGs. The total numbers of SFGs and QGs are 165,517 and 10,631, respectively. The $H_{160}$ magnitude at the 50\% completeness is $\sim 26.5$ mag for the photo-$z$ galaxies in deep CANDELS fields. The details of the completeness estimates and values are presented in \citet{2014arXiv1403.3689S}.

\subsection{Sample of LBGs at $z=4-10$ \\ in CANDELS, HUDF09/12, and HFF}\label{subsec_lbg}

The second sample consists of $10,454$ LBGs at $z=4-10$ made by Y. Harikane et al. (in preparation) in the CANDELS, the {\it Hubble} Ultra Deep Field 09+12 \citep[HUDF 09+12; ][]{2006AJ....132.1729B,2011ApJ...737...90B,2013ApJS..209....6I,2013ApJ...763L...7E} fields\footnote{http://archive.stsci.edu/prepds/xdf/}, and the parallel fields of Abell 2744 and MACS0416 in the {\it Hubble} Frontier Fields \citep[e.g., ][]{2014arXiv1405.0011C,2015ApJ...800...18A, 2014arXiv1409.1228O,2014arXiv1408.6903I}. The numbers of our LBGs are summarized in Table \ref{table_number_lbg}. These LBGs are selected with the color criteria, similar to those of \citet{2014arXiv1403.4295B}. We perform source detections by {\tt SExtractor} \citep{1996A&AS..117..393B} in coadded images constructed from bands of $Y_{098}Y_{105}J_{125}H_{160}$, $J_{125}H_{160}$, and $H_{160}$ for the $z\sim4-7$, $8$, and $10$ LBGs, respectively. The $JH_{140}$ band is included in the coadded image for the $z\sim7-8$ LBGs in the HUDF09+12 field. The flux measurements are carried out in \cite{1980ApJS...43..305K}-type apertures with a \citeauthor{1980ApJS...43..305K} parameter of $1.6$ whose diameter is determined in the $H_{160}$ band. In two-color diagrams, we select objects with a Lyman break, no extreme-red stellar continuum, and no detection in passbands bluer than the spectral drop. See Y. Harikane et al. (in preparation) for more details of the source detections and LBG selections. 

The $H_{160}$ magnitudes at the 50\% completeness is $\sim28$ mag for the LBGs in deep CANDELS fields \citep{2014arXiv1403.4295B}. The details of the completeness estimates and values are presented in Y. Harikane et al. (in preparation). 

Several previous studies on galaxy size have included a galaxy at $z\sim 12$ selected in photo-$z$ technique \citep{2013ApJ...763L...7E}. In this study, we do not use the galaxy at $z\sim 12$ because the redshift of the source is under debate \citep[e.g., ][]{2013ApJ...763L...7E,2013ApJ...765L...2B,2013ApJ...765L..16B,2013ApJ...773L..14C,2013ApJ...775...11P}.

\begin{figure*}[t!]
  \begin{center}
    \includegraphics[width=140mm]{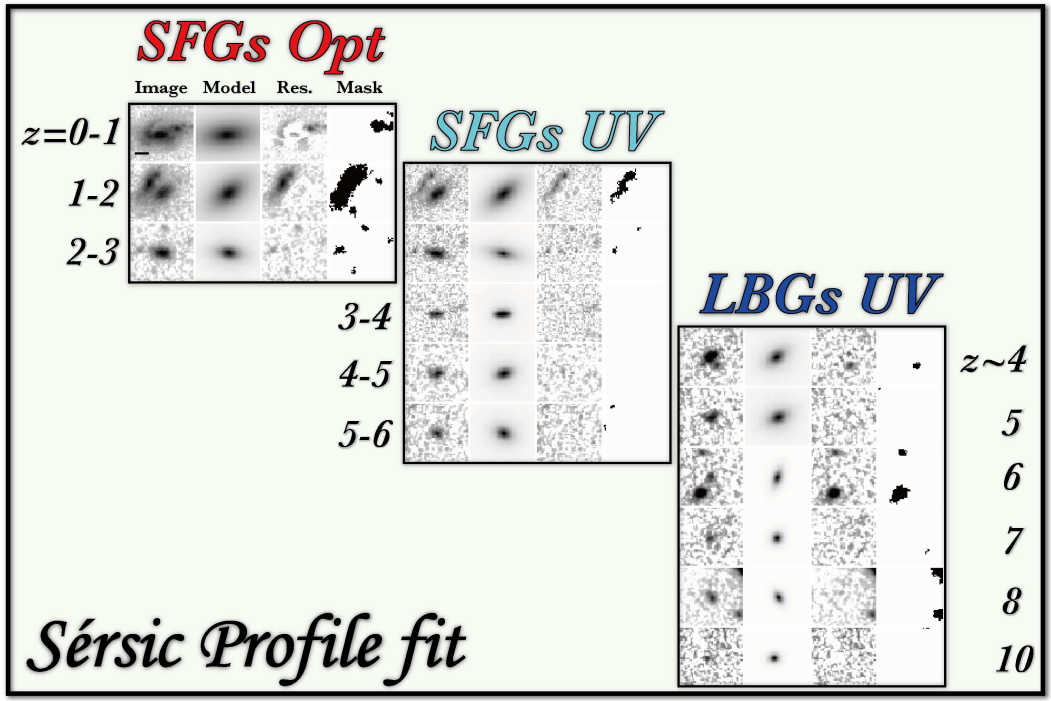}
  \end{center}
  \caption[]{{\footnotesize Examples of S\'ersic profile fitting results. The right, middle, and left panels indicate the results for the SFGs in the rest-frame optical and UV stellar continuum emission, and the LBGs, respectively. Each of 4-panel sets presents the original, the best-fit model, the residual, and the mask images, from left to right (see Section \ref{sec_analysis_size} for details) for one object. Each row, from top to bottom, denotes example galaxies from $z\sim0$ to $\sim10$. The SFGs at $z=1-2$ and $z=2-3$ are the same objects between the left and right panel sets that exhibit the images at the rest-frame optical and UV wavelengths, respectively. The black tick in the panel for the SFG at $z\sim0-1$ indicates the size of 1$^{\prime\prime}$. North is up and east is to the left. }}
  \label{fig_postage_stamps}
\end{figure*}

\begin{deluxetable*}{cccccccccc}
\setlength{\tabcolsep}{0.35cm} 
\tabletypesize{\scriptsize}
\tablecaption{Catalog of Photo-$z$ Galaxies with our Size Measurements}
\tablehead{\colhead{Catalog ID} & \colhead{$m_{\rm UV}$} & \colhead{$R_{e, {\rm major}}^{\rm UV}$} & \colhead{$n_{\rm UV}$} & \colhead{$q_{\rm UV}$} & \colhead{$m_{\rm Opt}$} & \colhead{$R_{e, {\rm major}}^{\rm Opt}$} & \colhead{$n_{\rm Opt}$} & \colhead{$q_{\rm Opt}$} & \colhead{flag} \\ 
\colhead{}& \colhead{[mag]}& \colhead{[arcsec]}& \colhead{}& \colhead{} & \colhead{[mag]} & \colhead{[arcsec]}& \colhead{} & \colhead{} & \colhead{}\\
\colhead{(1)}& \colhead{(2)}& \colhead{(3)}& \colhead{(4)}& \colhead{(5)}& \colhead{(6)}& \colhead{(7)} & \colhead{(8)}& \colhead{(9)}& \colhead{(10)}} 

\startdata

gds\_29 &  \nodata & \nodata & \nodata & \nodata & $23.70\pm0.021$ & $0.489\pm0.016$ &  $0.94\pm0.04$ &  $0.92\pm0.02$ & 0 \\
gds\_32 &  \nodata & \nodata & \nodata & \nodata &  $24.34\pm0.042$ & $0.164\pm0.013$ & $4.15\pm0.35$ & $0.77\pm0.03$ & 0 \\
gds\_59 &  \nodata & \nodata & \nodata & \nodata &  $25.31\pm0.029$ & $0.183\pm0.008$ & $0.65\pm0.09$ & $0.57\pm0.03$ & 0 \\
gds\_86 &  \nodata & \nodata & \nodata & \nodata &  $25.58\pm0.031$ & $0.174\pm0.008$ & $0.36\pm0.09$ & $0.73\pm0.04$ & 0 \\
gds\_122 &  \nodata & \nodata & \nodata & \nodata &  $25.70\pm0.051$ & $0.201\pm0.017$ & $0.66\pm0.14$ & $0.87\pm0.06$ & 0 \\
\nodata & \nodata & \nodata & \nodata & \nodata & \nodata & \nodata & \nodata & \nodata & \nodata 

\enddata

\tablecomments{A catalog of the photo-$z$ galaxies with $S/N\geq15$ and reliable outputs of {\tt GALFIT} fitting. Five example objects are shown here. Columns: (1) Catalog ID. The alphabetical characters represent the {\it HST} fields (``{\tt gds}"; GOODS-South, ``{\tt gdn}"; GOODSN-North, ``{\tt uds}"; UDS, ``{\tt aeg}"; AEGIS, ``{\tt cos}"; COSMOS). The numeric characters correspond to the ID number in the 3D-HST catalog \citep{2014arXiv1403.3689S}. (2) and (6) Total magnitude. (3) and (7) Effective radius along the major axis in arcseconds. (4) and (8) S\'ersic index. (5) and (9) Axis ratio. (10) Flag for the reliability of {\tt GALFIT} fitting. The values of {\tt 0} and {\tt 1} indicates reliable and unreliable measurements, respectively. (2)-(5) Measurements at $\lambda_{\rm int}^{\rm UV}=1500-3000$\,\AA. (7)-(9) Measurements at $\lambda_{\rm int}^{\rm Opt}=4500-8000$\,\AA. All measurement uncertainties are the half-width of the $68$\%-confidence interval.  \\ (The complete table is available in a machine-readable form in the online journal.) }
\label{table_catalog_photoz}
\end{deluxetable*}

\begin{deluxetable}{ccccc}
\setlength{\tabcolsep}{0.35cm} 
\tabletypesize{\scriptsize}
\tablecaption{Catalog of LBGs with our Size Measurements}
\tablehead{\colhead{Catalog ID} & \colhead{$m_{\rm UV}$} & \colhead{$R_{e, {\rm major}}^{\rm UV}$} & \colhead{$q_{\rm UV}$} & \colhead{flag} \\ 
\colhead{}& \colhead{[mag]}& \colhead{[arcsec]} & \colhead{} & \colhead{} \\
\colhead{(1)}& \colhead{(2)}& \colhead{(3)}& \colhead{(4)}& \colhead{(5)} } 

\startdata

z4\_gdsd\_10028 &  $26.42\pm0.05$ & $0.344\pm0.029$ & $0.16\pm0.03$ & 0 \\ 
z4\_gdsd\_10045 &  $25.23\pm0.02$ & $0.241\pm0.008$ & $0.57\pm0.02$ & 0 \\
z4\_gdsd\_10054 &  $26.58\pm0.03$ & $0.100\pm0.009$ & $0.47\pm0.08$ & 0 \\
z4\_gdsd\_10153 &  $27.60\pm0.07$ & $0.102\pm0.024$ & $0.43\pm0.21$ & 0 \\
z4\_gdsd\_10202 &  $26.33\pm0.04$ & $0.196\pm0.012$ & $0.38\pm0.05$ & 0 \\
\nodata & \nodata & \nodata & \nodata & \nodata  

\enddata

\tablecomments{A catalog of the LBGs with $S/N\geq15$ and reliable outputs of {\tt GALFIT} fitting. Five example objects are shown here. Columns: (1) Catalog ID in Y. Harikane et al. (in preparation) (2) Total magnitude. (3) Effective radius along the major axis in arcseconds. (4) Axis ratio. (5) Flag for the reliability of {\tt GALFIT} fitting. The values of {\tt 0} and {\tt 1} indicates reliable and unreliable measurements, respectively. The {\tt GALFIT} fitting is performed in the coadded {\it HST} images (see Section \ref{sec_analysis_size}). Note that S\'ersic indices are not listed due to fixed $n$ values in the {\tt GALFIT} fitting. All measurement uncertainties are the half-width of the $68$\%-confidence interval. \\ (The complete table is available in a machine-readable form in the online journal.) }
\label{table_catalog_lbg}
\end{deluxetable}

\begin{figure}[t!]
  \begin{center}
    \includegraphics[width=83mm]{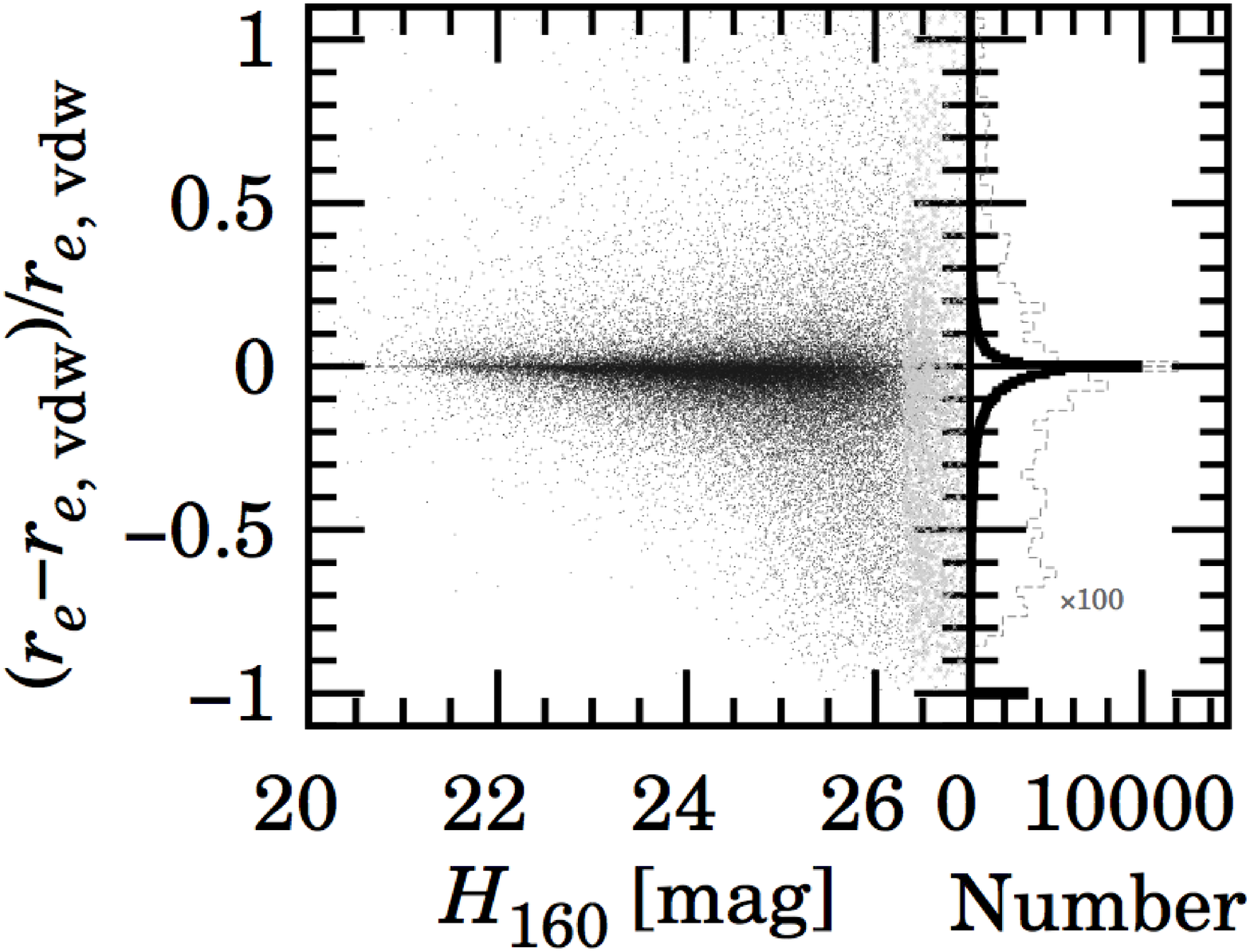}
  \end{center}
  \caption[]{{\footnotesize Comparison between the effective radii of $r_e$ and $r_{e, {\rm vdw}}$ measured by this study and \citet{2014ApJ...788...28V}, respectively, for objects with $S/N\geq15$ (black dots) and $<15$ (gray crosses at $H_{160}\gtrsim26.5$ mag). The right panel shows histograms for the number of the galaxies. The black and gray histograms denote objects with $S/N\geq15$ and $<15$, respectively. The number of objects with $S/N<15$ is multiplied by a factor of $100$ for clarity. }}
  \label{fig_mag_dre_hist}
\end{figure}

\subsection{Stellar Masses of Photo-$z$ Galaxies and LBGs}\label{subsec_stellarmass}

Some analyses and discussions in this work require $M_*$ of the photo-$z$ galaxies and the LBGs. For the photo-$z$ galaxies, we take $M_*$ values from \citet{2014arXiv1403.3689S}. For the LBGs, we derive stellar masses, adopting an empirical relation between UV magnitude $M_{\rm UV}$ and $M_*$. First, we calculate $M_{\rm UV}$ from the total magnitudes in the LBG detection images (Section \ref{sec_analysis_size}), assuming that the typical redshifts are $\langle z\rangle\sim 3.8, 4.9, 5.9, 6.8, 7.9$, and $10.4$. The stellar masses are obtained by converting their $M_{\rm UV}$ through the empirical \citeauthor{2011ApJ...735L..34G}'s relation \citep[see also, the updated result of ][]{2014ApJ...781...34G}, 

\begin{eqnarray}\label{eq_gonzalez}
\begin{split}
\log M_* &=& -39.6 + 1.7 \times \log L_{\rm 1500},  \\
 &=& -4.50-0.68\times M_{\rm UV},
\end{split}
\end{eqnarray}

%\begin{equation}\label{eq_gonzalez}
%\log M_* = -39.6 + 1.7 \times \log L_{\rm 1500}, 
%\end{equation}

\noindent where $L_{\rm 1500}$ is the luminosity at the rest-frame $1500$\,\AA. This empirical relation is derived under the assumptions similar to ours (the \citeauthor{1955ApJ...121..161S} IMF and no nebular emission lines included in SED). 

To test whether this empirical relation (eq. \ref{eq_gonzalez}) of $L_{\rm UV}$-$M_*$ is reliable and consistent with the $M_*$ estimates of the photo-$z$ galaxy sample, we compare this empirical relation with $M_{\rm UV}$-$M_*$ relations derived from the photo-$z$ galaxies.

We estimate $M_{\rm UV}$ from the absolute UV magnitudes at a wavelength of $2800$\,\AA\, from the photo-$z$ catalog, assuming the majority of star-forming galaxies have a flat UV spectrum of $f_\nu=$const. We present $M_{\rm UV}$-$M_*$ relations of the photo-$z$ galaxies in Figure \ref{fig_muv_mass_sfg}. The UV magnitude correlates well with $M_*$, suggesting the existence of the ``star-formation main sequence" \citep[e.g., ][]{2007ApJ...670..156D,2012ApJ...752...66L,2012ApJ...754L..29W,2014ApJ...791L..25S}. 

Figure \ref{fig_muv_mass_sfg} presents that the slopes of the relations appear to be flatter at a bright range of $M_{\rm UV}\lesssim-22$ than at a faint $M_{\rm UV}$ range. Similar flat slopes are reported by a large survey area of the CANDELS fields \citep{2009ApJ...697.1493S,2011ApJ...733...99L,2015ApJ...799..183S}. Because our LBGs used in this analysis have magnitudes of $M_{\rm UV}\geq-22$, we fit $\log{M_*} = a+b M_{\rm UV}$ to the $M_{\rm UV}$-$M_*$ relation at $M_{\rm UV}\geq-22$, where $a$ and $b$ are free parameters. The best-fit functions for the photo-$z$ galaxies are 

\begin{eqnarray}\label{eq_muv_mass_faint}
\begin{split}
\log{M_*} [M_\odot] &=& 1.46 -0.43\times M_{\rm UV} (z=0-1), \\
 &=& 0.82 -0.45\times M_{\rm UV} (z=1-2), \\
 &=& 1.12 -0.43\times M_{\rm UV} (z=2-3), \\
 &=& -0.22 -0.49\times M_{\rm UV} (z=3-4), \\
 &=& -2.10 -0.58\times M_{\rm UV} (z=4-5), \\
 &=& -2.45 -0.59\times M_{\rm UV} (z=5-6).
\end{split}
\end{eqnarray}

If we assume that the magnitudes of $1700$\,\AA\, to $1500$\,\AA\ are the same for typical LBGs with $f_\nu=$const, the slopes $b$ of $-0.58\pm0.02$ and $-0.59\pm0.03$ at $z\sim4-6$ roughly agree with that of eq. (\ref{eq_gonzalez}) (i.e., $b=-0.68\pm0.08$). We thus conclude that the empirical relation (eq. \ref{eq_gonzalez}) is reliable and consistent with the $M_*$ estimates of the photo-$z$ galaxy sample. Moreover, no strong evolution in the $M_{\rm UV}$-$M_*$ relation is found at $z\gtrsim4$ in eq. (\ref{eq_muv_mass_faint}) and Figure \ref{fig_muv_mass_sfg}. We use eq. (\ref{eq_gonzalez}) to estimate $M_*$ of our $z\gtrsim4$ LBGs.

%% OBSERVATION
%%%%%%%%%%%%%%%%%%%%%%%%%%%%%%%%%%%%%%%%%%%%%%%%
\section{Size Measurement}\label{sec_analysis_size}

In this section, we describe methods to measure galaxy sizes by using the high spatial resolution images of $\!${\it HST}. To minimize the effect of morphological {\it K}-correction, we use images of four bands, $V_{606}$ and $I_{814}$ on ACS\footnote{We make use of $z_{850}$ for GOODS-North that has not been taken with the $I_{814}$ band. }, and $J_{125}$ and $H_{160}$ on WFC3/IR. We select one of these bands whose entire passband is covered by the wavelength range of $\lambda_{\rm int}^{\rm UV}=1500-3000$\,\AA\, or $\lambda_{\rm int}^{\rm Opt}=4500-8000$\,\AA\, of each object. If two or more filter passbands meet this criterion, we chose a band that observes the shortest wavelength. Prior to the size measurements, we extract $18^{\prime\prime}\times18^{\prime\prime}$ cutout images from the $V_{606}I_{814}J_{125}H_{160}$ data at the position of each photo-$z$ galaxy and LBG. The size of cutout images is sufficiently large to investigate entire galaxy structures even for extended objects at $z\sim0-1$. We use coadded images of $Y_{098}Y_{105}J_{125}H_{160}$, $J_{125}H_{160}$, and $H_{160}$ constructed in Section \ref{subsec_lbg} for the $z\sim4-7$, $8$, and $10$ LBGs, respectively. The limiting magnitudes of these coadded images are summarized in Table \ref{table_hst_images}. 

We measure the galaxy size basically in the same manner as previous studies for high-$z$ LBGs \citep[e.g., ][]{2013ApJ...777..155O} based on the two-dimensional (2D) surface brightness (SB) profile fitting with the {\tt GALFIT} software \citep{2002AJ....124..266P,2010AJ....139.2097P}. We fit a single S\'ersic profile \citep{1963BAAA....6...41S,1968adga.book.....S} to the 2D SB distribution of each galaxy to obtain the half-light radius along the semi-major axis, $r_{e, {\rm major}}$. The $r_{e, {\rm major}}$ is converted to the ``circularized" radius, $r_e$, through $r_e\equiv a\sqrt{b/a}=r_{e, {\rm major}}\sqrt{q}$, where $a$, $b$, and $q$ are the major, minor axes, and axis ratio, respectively. Several authors studying $z\sim0-3$ galaxies claim that $r_{e, {\rm major}}$ should be used, because $r_{e, {\rm major}}$ does not depend strongly on the galaxy inclination \citep[e.g., ][]{2014ApJ...788...28V}. However, the circularized radius $r_e$ has widely been used in size measurements for faint and small high-$z$ sources \citep[e.g., ][]{2012ApJ...756L..12M,2013ApJ...777..155O,2014arXiv1406.1180H}. We here use the circularized radius $r_e$ in order to perform self-consistent size measurements and fair comparisons from $z\sim0$ to $\sim10$.  

We create {\it sigma} and {\it mask} images for estimating the fitting weight of individual pixels and masking neighboring objects of the main galaxy components, respectively. The sigma images are generated from the drizzle weight maps produced by the $\!${\it HST} data reduction \citep{2003hstc.conf..337K}. We also include the Poisson noise from the galaxy light to the sigma image \citep[e.g., ][]{2009ApJ...690.1866H,2012ApJS..203...24V}. The mask images are constructed from segmentation maps produced by {\tt SExtractor}. We identify neighboring objects with the {\tt SExtractor} detection parameters of {\tt DETECT\_MINAREA}$=5$ pixel, {\tt DETECT\_THRESH}$=2\sigma$, {\tt DETECT\_NTHRESH}$=16$, and {\tt DEBLEND\_MINCONT}$=0.0001$.

\begin{figure}[t!]
  \begin{center}
    \includegraphics[width=85mm]{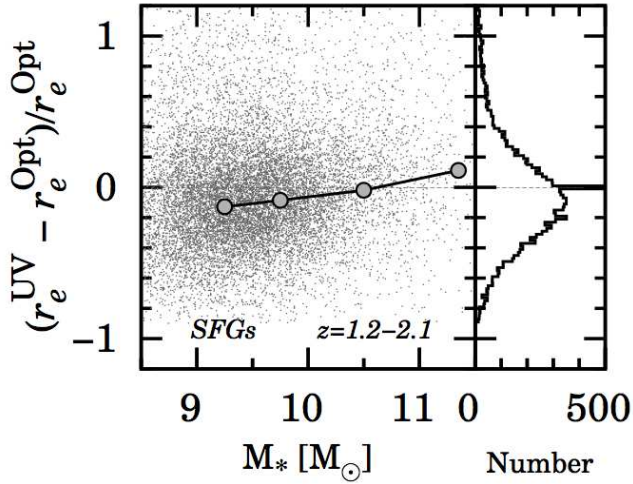}
  \end{center}
  \caption[]{{\footnotesize Difference between $r_e^{\rm UV}$ and $r_e^{\rm Opt}$ for the SFGs at $1.2\lesssim z\lesssim2.1$ (gray dots) as a function of stellar mass. The gray circles represent the median values of $(r_e^{\rm UV} - r_e^{\rm Opt})/r_e^{\rm Opt}$ in different stellar mass bins. The right panel shows histograms for the number of SFGs. }}
  \label{fig_mass_ruvropt_hist}
\end{figure}

We input initial parameters taken from the 3D-HST+CANDELS photometric catalogue \citep{2014arXiv1403.3689S} for the photo-$z$ galaxies. Specifically, the total magnitude $m$, axis ratio $q$, position angle $P.A.$, and half light radius $r_e$ of each galaxy are initial parameters that are written in the {\tt GALFIT} configuration file. The S\'ersic index $n$ is set to $n=1.5$ as an initial value for the photo-$z$ galaxies, while initial $n$ does not affect strongly fitting results \citep{2011ApJ...736...92Y,2012ApJ...761...19Y}. In fact, we change the initial parameters of S\'ersic index to $n=1$ and $3$, but still obtain similar best-fit $n$ values even with these different initial parameters. For the LBG sample, the initial parameters are taken from the results of {\tt SExtractor} photometry (Y. Harikane in preparation). The S\'ersic index for LBGs is fixed to $1.5$ for reliable fitting for faint and small high-$z$ sources. This fixed S\'ersic index is justified by the evolution of $n$ in SFGs, as demonstrated in Section \ref{subsec_results_z_n}. To obtain $r_e, n$, and $q$, we allow the parameters to vary in the ranges, $\Delta m<3$ mag, $0.3<r_e<400$ pixels, $0.2<n<8$, $0.0001<q<1$, $\Delta x<2$ pixel, and $\Delta y<2$ pixel, which are quite similar to those of \citet{2012ApJS..203...24V}. We discard objects whose one or more fitting parameters reach the limit of the parameter ranges (e.g., $r_e=400$). The PSF models of the $\!${\it HST} images are provided from the 3D-HST project \citep{2014arXiv1403.3689S}.

\begin{deluxetable*}{ccccccc}
\setlength{\tabcolsep}{0.35cm} 
\tabletypesize{\scriptsize}
\tablecaption{Summary of the Best-Fit Size Growth Rates}
\tablehead{\colhead{Data points} & \colhead{Sample} & \colhead{$L_{\rm UV}/L_{z=3}^*$} & \colhead{$B_z$} & \colhead{$\beta_z$} & \colhead{$B_H$} & \colhead{$\beta_H$} \\ 
\colhead{}& \colhead{}& \colhead{}& \colhead{[kpc]}& \colhead{}& \colhead{[kpc]}& \colhead{}\\
\colhead{(1)}& \colhead{(2)}& \colhead{(3)}& \colhead{(4)}& \colhead{(5)}& \colhead{(6)}& \colhead{(7)}} 

\startdata
Median & All & 1-10 & $4.78\pm0.68$ & $-0.84\pm0.11$ & $3.80\pm0.40$ & $-0.62\pm0.08$ \\
 & & 0.3-1 & $5.45\pm0.31$ & $-1.10\pm0.06$ & $4.33\pm0.17$ & $-0.86\pm0.04$ \\
 & & 0.12-0.3 & $4.44\pm0.19$ & $-1.22\pm0.05$ & $3.46\pm0.13$ & $-0.97\pm0.05$ \\
 & w/o $r_e^{\rm Opt}$ & 1-10 & $4.05\pm0.59$ & $-0.78\pm0.08$ & $3.09\pm0.36$ & $-0.56\pm0.06$ \\
 & & 0.3-1 & $5.21\pm0.28$ & $-1.15\pm0.07$ & $3.54\pm0.29$ & $-0.80\pm0.05$ \\
 & & 0.12-0.3 &  $3.54\pm0.58$ & $-1.11\pm0.11$ & $2.45\pm0.32$ & $-0.78\pm0.08$ \\ \hline

Average & All & 1-10 & $5.80\pm0.65$ & $-0.79\pm0.10$ &$4.91\pm0.42$ & $-0.61\pm0.07$ \\
 & & 0.3-1 & $5.85\pm0.33$ & $-0.95\pm0.07$ &$4.83\pm0.20$ & $-0.74\pm0.04$ \\
 & & 0.12-0.3 &  $5.52\pm0.43$ & $-1.17\pm0.07$ &$4.29\pm0.27$ & $-0.87\pm0.05$ \\
 & w/o $r_e^{\rm Opt}$ & 1-10 & $11.3\pm4.44$ & $-1.22\pm0.25$ & $7.48\pm2.37$ & $-0.85\pm0.18$ \\ 
 & & 0.3-1 & $10.9\pm2.94$ & $-1.36\pm0.18$ & $6.90\pm1.50$ & $-0.95\pm0.13$ \\ 
 & & 0.12-0.3 & $6.82\pm2.25$ & $-1.31\pm0.20$ & $4.37\pm1.18$ & $-0.91\pm0.14$ \\ \hline

Mode  & All & 1-10 & $4.00\pm0.49$ & $-0.78\pm0.08$ & $3.07\pm0.30$ & $-0.55\pm0.57$ \\
 & & 0.3-1 & $4.45\pm0.89$ & $-1.26\pm0.17$ & $2.97\pm0.45$ & $-0.89\pm0.12$ \\
 & & 0.12-0.3 & $3.28\pm0.18$ & $-1.23\pm0.07$ & $2.56\pm0.11$ & $-1.00\pm0.05$ \\
 & w/o $r_e^{\rm Opt}$ & 1-10 & $10.9\pm3.94$ & $-1.14\pm0.25$ & $7.45\pm2.14$ & $-0.80\pm0.17$ \\ 
 & & 0.3-1 & $3.00\pm0.19$ & $-1.01\pm0.05$ & $2.15\pm0.10$ & $-0.71\pm0.03$ \\ 
 & & 0.12-0.3\tablenotemark{a} & \nodata & \nodata & \nodata & \nodata  

\enddata

\tablecomments{Columns: (1) Statistics of $r_e$. (2) Sample used in the fits for the size evolution. ``All" denotes to use all samples in a $L_{\rm UV}$ bin. ``w/o $r_e^{\rm Opt}$" represents to exclude the data points of $r_e^{\rm Opt}$. (3) Bins of $L_{\rm UV}$ in units of $L_{z=3}^*$. (4) $B_z$ of $B_z (1+z)^{\beta_z}$. (5) $\beta_z$ of $B_z (1+z)^{\beta_z}$. (6) $B_H$ of $B_H h(z)^{\beta_H}$, where $h(z) \equiv H(z) / H_0 = \sqrt{\Omega_m(1+z)^3 + \Omega_\Lambda}$. (7) $\beta_H$ of $B_H h(z)^{\beta_H}$.}
\tablenotetext{a}{The $\chi^2$ minimization is not converged.}
\label{table_fits}
\end{deluxetable*}

\begin{figure*}[t!]
  \begin{center}
    \includegraphics[width=150mm]{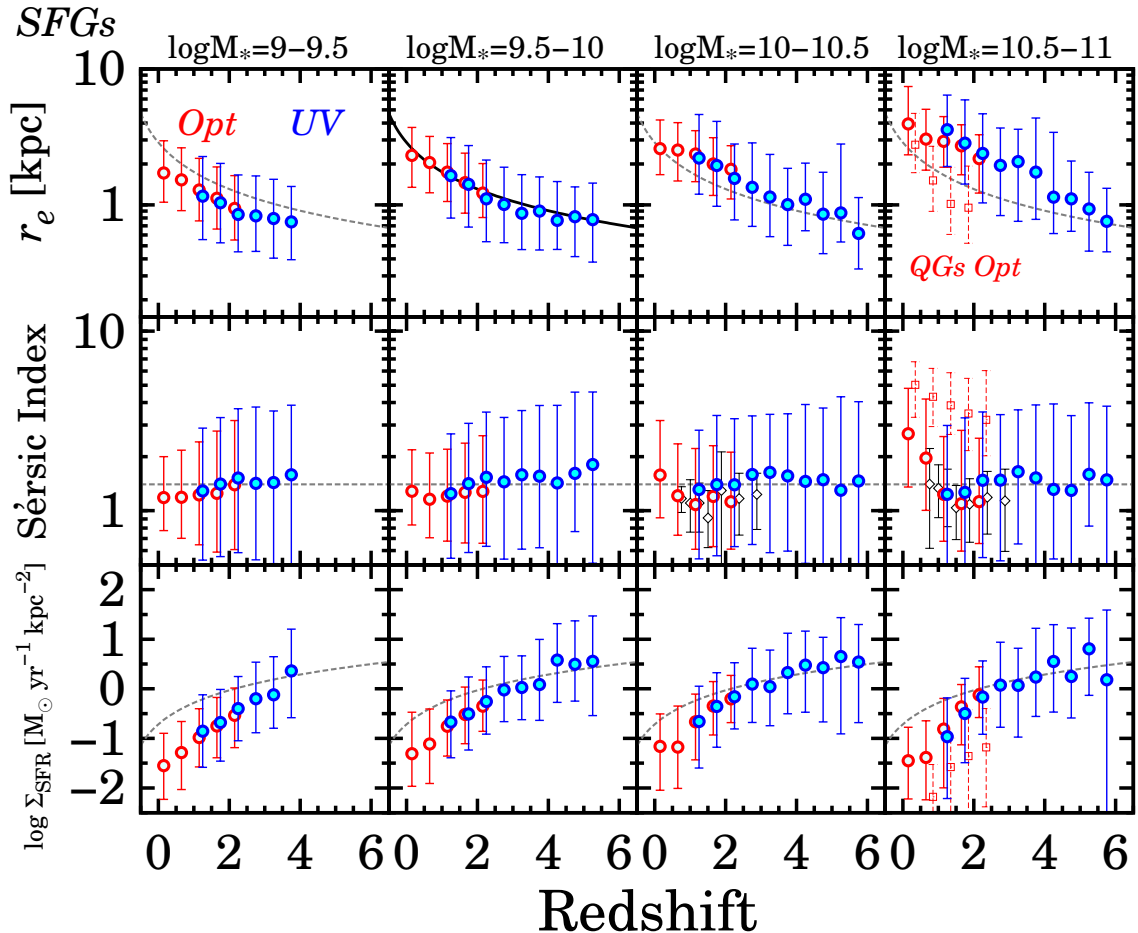}
  \end{center}
  \caption[]{{\footnotesize Redshift evolution of size-relevant quantities for the photo-$z$ galaxies in different stellar mass bins ($\log{M_*}=9-9.5, 9.5-10, 10-10.5, $ and $10.5-11$ M$_\odot$ from left to right). From top to bottom, the panels show the effective radius $r_e$, S\'ersic index $n$, and SFR SD $\Sigma_{\rm SFG}$. The blue and red circles indicate median values of $r_e^{\rm UV}$ and $r_e^{\rm Opt}$, respectively, for the SFGs. The open squares in the rightmost panels present median values of $r_e^{\rm Opt}$ for the QGs with $\log{M_*}=10.5-11\, M_\odot$. The error bars denote the 16th and 84th percentiles of the data point distribution. In the top panels, the best-fit $r_e$ curve in the bin of $\log M_*=9.5-10\, M_\odot$ is plotted for reference with the solid and dashed lines. The best-fit $\beta_z$ and $\beta_H$ values are $-0.72\pm0.04$ and $-0.60\pm0.02$, respectively. The dashed gray lines in the bottom panels represent the $\Sigma_{\rm SFG}$ evolution calculated with SFR$=10\,M_\odot/$yr and the best-fit $r_e$ curve. The horizontal lines in the second-top panels denote a weighted mean of $\left< n\right>=1.4$. In the second-right and right-most panels of the Sersic index plots, the open diamonds denote measurements of S\'ersic index for SFGs in \citet{2014ApJ...785...18M}. The S\'ersic index estimates of \citet{2014ApJ...785...18M} are comparable to ours. In the panels of low mass $\log\, M_*=9-9.5\, M_\odot$, the quantities for high redshifts are not plotted, due to their poor statistics.}}
  \label{fig_sfg_tile}
\end{figure*}

\begin{deluxetable*}{cccccc}
\setlength{\tabcolsep}{0.35cm} 
\tabletypesize{\scriptsize}
\tablecaption{Summary of the LBG Size Growth Rates from the Previous Studies}
\tablehead{\colhead{Reference} & \colhead{Number} & \colhead{Redshift Range} & \colhead{$\beta_z$ of $(1+z)^{\beta_z}$} & \colhead{Statistics} & \colhead{Size Measurements} \\
\colhead{(1)}& \colhead{(2)}& \colhead{(3)}& \colhead{(4)} &  \colhead{(5)}& \colhead{(6)}} 

\startdata
\citet{2004ApJ...611L...1B} & \nodata $(2929)$ & $2-6$ & $-1.05\pm0.21$ & Average & {\tt SExtractor}\\
\citet{2004ApJ...600L.107F} & \nodata $(773)$ & $2-5$ & $\sim-1.5$ & Average & {\tt SExtractor}\\
\citet{2006ApJ...652..963R} & 1333 $(4694)$ & $3-5$ & \nodata & \nodata & {\tt GALFIT}\\
\citet{2008AJ....135..156H} & 61 $(61)$ & $3-6$ & $\sim-1.5$ & Average & {\tt SExtractor}\\
\citet{2009MNRAS.397..208C} & 583 $(583)$ & $4-6$ & \nodata & \nodata & {\tt SExtractor}\\
\citet{2010ApJ...709L..21O} & 21 $(21)$ & $7-8$ & $-1.12\pm0.17$ & Average & {\tt SExtractor}, {\tt GALFIT}\\
%\citet{2011ApJ...727....5M} & \nodata $(151)$ & $0-3$ & $-1.11\pm0.13$ & Median & {\tt GALFIT}\\
\citet{2012AA...547A..51G} & \nodata $(153)$ & $7$ & \nodata & \nodata & {\tt SExtractor}\\
\citet{2012ApJ...756L..12M} & \nodata $(218)$ & $4-7$ & $-1.20\pm0.11$ & Median & {\tt GALFIT}\\
\citet{2013ApJ...765...68H} & 1012 $(1356)$ & $4-5$ & $\sim-1$ & Mode  & {\tt SExtractor}, {\tt GALFIT}\\
\citet{2013ApJ...777..155O} & 15 $(81)$ & $7-10$ & $-1.30^{+0.12}_{-0.14}$ & Average & {\tt GALFIT}\\
\citet{2014arXiv1409.1832C} & 1318 $(3738)$ & $4-9$ & $-0.31\pm0.26$ & Mode & {\tt SExtractor}\\
\citet{2014arXiv1406.1180H} & 8 $(8)$ & $9-10$ & $-1.0\pm0.1$ & Average & {\tt GALFIT}\\
\citet{2014arXiv1410.1535K}\tablenotemark{a} & 39 $(39)$ & $6-8$ & $-1.24\pm0.1$ & Average & {\tt glafic}\\ \hline
This work & 4993 $(10454)$ & $4-10$ & $-1.10\pm0.06$ & Median & {\tt GALFIT}\\
 & & & $-0.95\pm0.07$ & Average & {\tt GALFIT}\\
 & & & $-1.26\pm0.17$ & Mode  & {\tt GALFIT}\\
incl. Photo-$z$ SFGs & 89880 $(312722)$\tablenotemark{b} & $0-6$ &  &  & 

\enddata

\tablecomments{Columns: (1) Reference. (2) Number of galaxies whose size is measured in the reference. The values in parentheses are the number of galaxies in parent sample. (3) Redshift range for size measurements of LBGs. (4) Best-fit $\beta_z$ of $(1+z)^\beta_z$ for a bright ($L_{\rm UV}\sim0.3-1\,L_{z=3}^*$) galaxy sample. (5) Statistics for deriving a representative $r_e$ at a redshift. ``Mode " corresponds to the peak of size distribution derived by the fitting with a log-normal function (Equation \ref{eq_lognormal}). (6) Method or software to measure galaxy sizes.}
\tablenotetext{a}{Sample galaxies are selected in a field of galaxy cluster. This study
corrects for the gravitational lensing effects of magnification and shear with their mass model.}
\tablenotetext{b}{The value is a total number of SFGs whose sizes are well measured in $r_e^{\rm Opt}$ and $r_e^{\rm UV}$. See Table \ref{table_number_sfg}. }
\label{table_previous_studies}
\end{deluxetable*}

We have analyzed the photo-$z$ galaxies and LBGs shown in Sections \ref{subsec_3dhst} and \ref{subsec_lbg}. As we discuss below, sizes of faint galaxies are poorly determined. We thus choose photo-$z$ galaxies and LBGs whose sources have a signal-to-noise ratio (S/N) greater than $15$. This $S/N$threshold is determined by Monte Carlo simulations for faint and small high-$z$ sources \citep[e.g., ][]{2012ApJS..203...24V, 2013ApJ...777..155O}. Tables \ref{table_number_sfg} and \ref{table_number_lbg} summarize the number of photo-$z$ galaxies and LBGs, respectively, that are analyzed in our study. The object numbers in our size analysis are 142,273 (9,767) in $V_{606}$, 136,493 (10,118) in $I_{814}$, 139,308 (10,845) in $J_{125}$, and 147,204 (11,297) in $H_{160}$, for the SFGs (QGs) of the photo-$z$ sample, and $7,233$ for the LBGs. The total numbers of SFGs (QGs) that are well fit in the optical and UV stellar continuum emission are $59,115$ (4,234) at $z\sim0-3$ and $30,765$ (799) at $z\sim1-6$, respectively, while sizes of $4,993$ LBGs are securely measured. Tables \ref{table_catalog_photoz} and \ref{table_catalog_lbg} show the size measurements given by our structural analysis for the photo-$z$ galaxies and the LBGs, respectively. Figure \ref{fig_postage_stamps} presents example images of the fitting results, demonstrating that our size measurements are well performed. 

Note that clumpy structures are masked in the fitting, as indicated in the mask panels of Figure \ref{fig_postage_stamps}. This masking procedure is included in our analyses, because a single S\'ersic profile fitting is not reliable for galaxies with the clumpy structures. Moreover, the number of well-fit galaxies decreases, if no masking is applied. Nevertheless, we examine whether the masking procedures change our conclusions, and find that the $r_e$ measurements are statistically comparable in galaxies with and without masking. The fraction of galaxies with the clumpy structures ranges from $\sim30\%$ at $z\sim1$ to $\sim50\%$ at $z\sim2$. This study only addresses galaxies' major stellar components. The detailed analyses and the results of clumpy stellar sub-components are presented in the paper II.

%Figure \ref{fig_galfit} shows examples of original $I_{814}$ images, the best-fit S\'ersic profiles, and their residual images. As shown in Figure \ref{fig_galfit}, the {\tt GALFIT} fits well the stellar-continuum emission to S\'ersic profiles. 

\citet{2012ApJS..203...24V,2014ApJ...788...28V} obtain their $r_e$ values in the $J_{125}$, $JH_{140}$, and $H_{160}$ bands for all the 3D-HST+CANDLES galaxies using the {\tt GALAPAGOS} software \citep{2012MNRAS.422..449B} which is a wrapper of {\tt SExtractor} and {\tt GALFIT} for morphological analyses. Several morphological studies have utilized {\tt GALAPAGOS} allowing for the simultaneous determination of both the structural parameters and the background flux level for multi-objects. In Figure \ref{fig_mag_dre_hist}, we compare $r_e$ measurements of ours with those of \citet{2014ApJ...788...28V} estimated with {\tt GALAPAGOS}. We find that our $r_e$ values are in good agreement with those obtained by \citet{2014ApJ...788...28V}. We also find that faint galaxies with $S/N<15$ are significantly scattered in Figure \ref{fig_mag_dre_hist}. This confirms that the threshold of $S/N\geq15$ is important for secure size measurements. 

%The figure also shows no systematic discrepancies between the two $r_e$, ensuring the assessment of galaxy size evolution including the ACS bands. 

\begin{deluxetable}{ccccccc}
\setlength{\tabcolsep}{0.35cm} 
\tabletypesize{\scriptsize}
\tablecaption{Size-Luminosity Relation at $z=0-8$}
\tablehead{\colhead{$M_{\rm UV}$} & \colhead{$r_e^{\rm Opt}$} & \colhead{$M_{\rm UV}$} & \colhead{$r_e^{\rm UV}$} &\colhead{$M_{\rm UV}$} & \colhead{$r_e^{\rm UV}$}\\
\colhead{[mag]}& \colhead{[kpc]}& \colhead{[mag]}& \colhead{[kpc]}&\colhead{[mag]}& \colhead{[kpc]}\\
\colhead{(1)}& \colhead{(2)}& \colhead{(3)}& \colhead{(4)} &  \colhead{(5)}& \colhead{(6)} } 

\startdata

\multicolumn{2}{c}{$z=0-1$ SFGs} & \multicolumn{2}{c}{$z=1-2$ SFGs} & \multicolumn{2}{c}{$z\sim4$ LBGs} \\
$-21.0$ & $3.896^{+4.513}_{-2.273}$ & $-21.0$ & $2.090^{+2.847}_{-1.139}$ & $-23.0$ & $1.536^{+1.041}_{-0.878}$ \\
$-20.0$ & $3.380^{+2.036}_{-1.497}$ & $-20.0$ & $1.794^{+2.418}_{-0.962}$ & $-22.0$ & $1.223^{+1.100}_{-0.600}$ \\
$-19.0$ & $2.504^{+1.566}_{-1.045}$ & $-19.0$ & $1.232^{+1.464}_{-0.634}$ & $-21.0$ & $1.058^{+0.841}_{-0.529}$ \\
$-18.0$ & $1.837^{+1.234}_{-0.810}$ & $-18.0$ & $0.940^{+1.056}_{-0.477}$ & $-20.0$ & $0.733^{+0.638}_{-0.325}$ \\
$-17.0$ & $1.340^{+0.983}_{-0.570}$ & $-17.0$ & $0.723^{+1.016}_{-0.352}$ & $-19.0$ & $0.589^{+0.510}_{-0.265}$ \\
$-16.0$ & $1.074^{+0.791}_{-0.467}$ & $-16.0$ & $0.606^{+1.108}_{-0.404}$ & $-18.0$ & $0.509^{+0.468}_{-0.204}$ \\
$-15.0$ & $0.854^{+0.680}_{-0.359}$ & $-15.0$ & $0.461^{+0.398}_{-0.309}$ & $-17.0$ & $0.438^{+0.325}_{-0.247}$ \\
\multicolumn{2}{c}{$z=1-2$ SFGs} & \multicolumn{2}{c}{$z=2-3$ SFGs} & \multicolumn{2}{c}{$z\sim5$ LBGs} \\
$-22.0$ & $3.035^{+0.237}_{-0.448}$ & $-22.0$ & $1.428^{+1.602}_{-0.632}$ & $-22.0$ & $1.025^{+1.044}_{-0.352}$ \\
$-21.0$ & $1.958^{+2.304}_{-1.026}$ & $-21.0$ & $1.443^{+1.664}_{-0.727}$ & $-21.0$ & $0.788^{+0.686}_{-0.359}$ \\
$-20.0$ & $1.982^{+1.414}_{-0.734}$ & $-20.0$ & $1.076^{+1.177}_{-0.545}$ & $-20.0$ & $0.595^{+0.393}_{-0.287}$ \\
$-19.0$ & $1.360^{+1.071}_{-0.589}$ & $-19.0$ & $0.813^{+0.787}_{-0.382}$ & $-19.0$ & $0.519^{+0.496}_{-0.232}$ \\
$-18.0$ & $1.113^{+0.817}_{-0.399}$ & $-18.0$ & $0.685^{+0.749}_{-0.333}$ & $-18.0$ & $0.506^{+0.426}_{-0.284}$ \\
$-17.0$ & $1.028^{+0.893}_{-0.451}$ & $-17.0$ & $0.509^{+0.573}_{-0.263}$ & $-17.0$ & $0.356^{+0.095}_{-0.076}$ \\
$-16.0$ & $0.937^{+1.017}_{-0.378}$ & \multicolumn{2}{c}{$z=3-4$ SFGs} & \multicolumn{2}{c}{$z\sim6$ LBGs} \\
\multicolumn{2}{c}{$z=2-3$ SFGs} & $-22.0$ & $1.473^{+1.687}_{-0.705}$ & $-22.0$ & $1.053^{+0.841}_{-0.696}$ \\
$-21.0$ & $2.878^{+3.607}_{-0.908}$ & $-21.0$ & $1.054^{+1.154}_{-0.495}$ & $-21.0$ & $0.635^{+0.717}_{-0.274}$ \\
$-20.0$ & $1.253^{+0.488}_{-0.098}$ & $-20.0$ & $0.778^{+0.765}_{-0.374}$ & $-20.0$ & $0.565^{+0.400}_{-0.287}$ \\
$-19.0$ & $1.066^{+0.543}_{-0.710}$ & $-19.0$ & $0.620^{+0.538}_{-0.278}$ & $-19.0$ & $0.584^{+0.424}_{-0.327}$ \\
$-18.0$ & $1.625^{+9.075}_{-0.997}$ & $-18.0$ & $0.572^{+0.577}_{-0.269}$ & $-18.0$ & $0.371^{+0.222}_{-0.198}$ \\
\nodata & \nodata & \multicolumn{2}{c}{$z=4-5$ SFGs} & \multicolumn{2}{c}{$z\sim7$ LBGs} \\
\nodata & \nodata & $-22.0$ & $1.081^{+1.082}_{-0.436}$ & $-21.0$ & $0.737^{+0.320}_{-0.421}$ \\
\nodata & \nodata & $-21.0$ & $0.892^{+0.769}_{-0.427}$ & $-20.0$ & $0.489^{+0.956}_{-0.268}$ \\
\nodata & \nodata & $-20.0$ & $0.708^{+0.703}_{-0.354}$ & $-19.0$ & $0.467^{+0.572}_{-0.207}$ \\
\nodata & \nodata & $-19.0$ & $0.444^{+0.362}_{-0.302}$ & \multicolumn{2}{c}{$z\sim8$ LBGs} \\
\nodata & \nodata & \multicolumn{2}{c}{$z=5-6$ SFGs} & $-21.0$ & $0.419^{+1.981}_{-0.262}$ \\
\nodata & \nodata & $-22.0$ & $0.975^{+3.757}_{-0.425}$ & $-20.0$ & $0.425^{+1.331}_{-0.173}$ \\
\nodata & \nodata & $-21.0$ & $0.716^{+0.674}_{-0.286}$ & $-19.0$ & $0.243^{+0.225}_{-0.068}$ \\
\nodata & \nodata & $-20.0$ & $0.678^{+0.756}_{-0.328}$  & $-18.0$ & $0.356^{+1.194}_{-0.218}$

\enddata

\tablecomments{Columns: (1) (3) (5) UV magnitude. (2) (4) (6) Median effective radius at the rest-frame optical or UV wavelength. The lower and upper limits indicate the 16th and 84th percentiles of the $r_e$ distribution, respectively.  }
\label{table_muv_re}
\end{deluxetable}

%% RESULTS
%%%%%%%%%%%%%%%%%%%%%%%%%%%%%%%%%%%%%%%%%%%%%%%%
%%%%%%%%%%%%%%%%%%%%%%%%%%%%%%%%%%%%%%%%%%%%%%%%
\section{$K$-Correction, Statistical Choice, and Selection Bias}\label{sec_bias}

%% Morphological K-Correction
%%%%%%%%%%%%%%%%%%%%%%%%%%%%%%%%%%%%%%%%%%%%%%%%
\subsection{Effect of Morphological K-Correction}\label{subsec_bias_kcorr}

We investigate the effects of morphological {\it K}-correction in our size measurements, comparing our $r_e$ at different wavelengths. Because the $\!${\it HST} imaging data covers up to $H_{160}$ band, we can study the rest-frame UV morphology for galaxies at $z\gtrsim3$. Understanding the effects of morphological {\it K}-correction is considerably important to evaluate the size evolution of star-forming galaxies over a wide redshift range of $0\lesssim z\lesssim10$. The sizes in the rest-frame UV and optical stellar continuum emission, $r_e^{\rm UV}$ and $r_e^{\rm Opt}$, tracing different stellar population would yield a large difference in $r_e$. Here we make a comparison between $r_e^{\rm UV}$ and $r_e^{\rm Opt}$ of the SFGs at $1.2\lesssim z\lesssim2.1$ where the both radii can be measured with the $\!${\it HST} data. 

Figure \ref{fig_mass_ruvropt_hist} shows the differences between $r_e^{\rm UV}$ and $r_e^{\rm Opt}$ of the SFGs as a function of stellar mass. Although we find a large scatter, the median values of $(r_e^{\rm UV}-r_e^{\rm Opt})/r_e^{\rm Opt}$ are less than $20$ \% in all stellar mass bins. This indicates that the differences of statistical $r_e$ measurements are small for star-forming galaxies with $\log{\rm M}_*=9-11$ ${\rm M}_\odot$ at $z\sim1-2$. 

Similarly, \citet{2014ApJ...788...28V} have found that $r_{e, {\rm major}}$ is typically smaller in redder bands for SFGs at $z\sim0-2$ \citep[see also, e.g., ][]{2011ApJ...735L..22S, 2012ApJ...753..114W}. This trend is more significant in more massive SFGs. The smaller size in redder bands would be interpreted as heavier dust attenuation in the galactic central regions in bluer bands \citep[e.g., ][]{2012MNRAS.421.1007K} and/or the inside-out disk formation \citep[e.g., ][]{2009ApJ...694..396B,2009ApJ...697.1290B,2009ApJ...699L.178N, 2012ApJ...747L..28N, 2013ApJ...766...15P}. We confirm the wavelength dependence even in our $r_e$ in the most massive M$_*$ bin as shown in Figure \ref{fig_mass_ruvropt_hist}. \citet{2014ApJ...788...28V} have parametrized the wavelength dependence of $r_{e, {\rm major}}$ as a function of redshift and stellar mass. Following the formula, the size difference fraction $(r_e^{\rm UV}-r_e^{\rm Opt})/r_e^{\rm Opt}$ is calculated to be $\sim30$\% for $z\sim2$ galaxies with $\log{M_*}=11$ M$_\odot$. 

Note that the difference of stellar population becomes smaller at $z>2$ than $z\sim 1-2$, because the short cosmic age of $z>2$ provides a smaller stellar-age difference and a less metal enrichment than that of $z\sim 1-2$. This agreement of $r_e^{\rm UV}$ and $r_e^{\rm Opt}$ suggests that the statistical $r_e^{\rm UV}$ values represent the typical sizes of stellar-component distribution for star-forming galaxies of SFGs and LBGs at $z\gtrsim3$ with a small systematic uncertainty of $\lesssim 30$ \%\footnote{In Figure \ref{fig_mass_ruvropt_hist}, we find that the scatters of $(r_e^{\rm UV}-r_e^{\rm Opt})/r_e^{\rm Opt}$ are comparably large in high and low-mass galaxies. Because the scatters originated from statistical errors should be smaller in the high-mass galaxies than the low-mass galaxies, the scatters of the high-mass galaxies are probably not dominated by statistical errors but intrinsic $r_e$ differences.}. 

We examine the effect of morphological {\it K}-correction in more detail by investigating evolutionary trends of $r_e$ and size-relevant quantities in the rest-frame optical and UV emission for the photo-$z$ galaxies at $z\sim0-6$. Figure \ref{fig_sfg_tile} presents redshift evolution of $r_e$, $n$, and star-formation rate surface density (SFR SD), $\Sigma_{\rm SFR}$. The SFR SD is derived in the effective radius, and calculated by

\begin{equation}\label{eq_sfrsd}
\Sigma_{\rm SFR}\,[M_\odot\,{\rm yr}^{-1}\,{\rm kpc}^{-2}]= \frac{{\rm SFR}/2}{\pi r_e^2}, 
\end{equation}

\noindent where a factor of $1/2$ corrects for the SFR value which is derived from the total magnitudes. For the photo-$z$ galaxies, we use SFRs taken from the catalog of \citet{2014arXiv1403.3689S}. For LBGs, we compute SFRs from $L_{\rm UV}$ using the relation of \citet{1998ARA&A..36..189K},

\begin{equation}\label{eq_sfr}
{\rm SFR}\,[M_\odot\,{\rm yr}^{-1}]= 1.4 \times 10^{-28} L_\nu\,[{\rm erg s}^{-1}\,{\rm Hz}^{-1}]. 
\end{equation}

\citet{2014ApJ...788...28V} have already examined the $r_e$ evolution at $5000$\,\AA\, in the rest-frame for galaxies at $0\lesssim z\lesssim3$ in the 3D-HST+CANDELS sample. In our study, we extend this analysis of $0\lesssim z\lesssim3$ to $z\gtrsim 4$, using the photo-$z$ galaxies and the LBGs.

In Figure \ref{fig_sfg_tile}, median values of these quantities are in good agreement between the measurements in the rest-frame optical and UV emission of the SFGs at $2\lesssim z\lesssim3$. Additionally, the evolutionary tracks at $z\lesssim3$ smoothly connect with those at $z\gtrsim3$. We also find no strong dependence of these evolutionary trends on the stellar mass. These agreements confirm a small effect of morphological {\it K}-correction in the median $r_e$ values.

%% Statistical Difference
%%%%%%%%%%%%%%%%%%%%%%%%%%%%%%%%%%%%%%%%%%%%%%%%
\subsection{Statistical Difference and Selection Bias}\label{subsec_bias_statistics}

\begin{figure}[t!]
  \begin{center}
    \includegraphics[width=87mm]{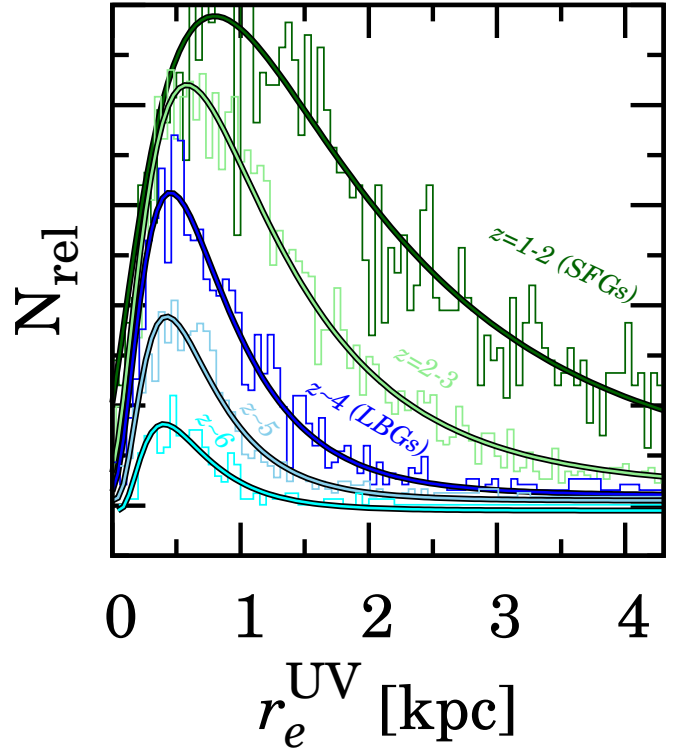}
  \end{center}
  \caption[]{{\footnotesize Distribution of $r_e^{\rm UV}$ for the SFGs and the LBGs at $z\sim1-6$ in the bin of $L_{\rm UV}=0.3-1\,L_{z=3}^*$. The histograms and the curves show the $r_e^{\rm UV}$ distributions and the best-fit log-normal functions, respectively, for the SFGs at $z=1-2$ (green) and $2-3$ (light-green) and the LBGs at $z\sim4$ (blue), $5$ (light-blue), and $6$ (cyan). The y-axis is arbitrary. The histograms and curves are slightly shifted along $x$- and $y$- axes for clarity. The shifted values are $\Delta r_e=-0.25, -0.12, -0.09, -0.04$, and $0$ kpc for $z=1-2$ and $z=2-3$ SFGs, and $z\sim4$, $z\sim5$, and $z\sim6$ LBGs, respectively. Although these choices of the shifts moderately cancel out the trend of the $r_e$ evolution, the $r_e$ decrease towards high-$z$ is still clearly found. }}
  \label{fig_re_hist_large}
\end{figure}

\begin{figure*}[t!]
  \begin{center}
    \includegraphics[height=220mm]{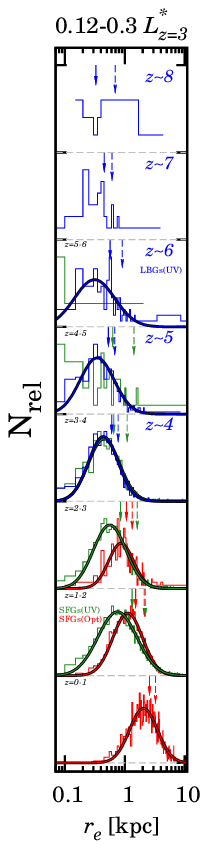}
    \includegraphics[height=220mm]{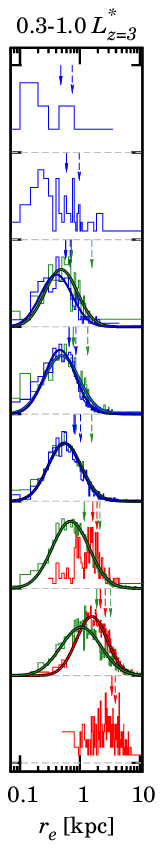}
    \includegraphics[height=220mm]{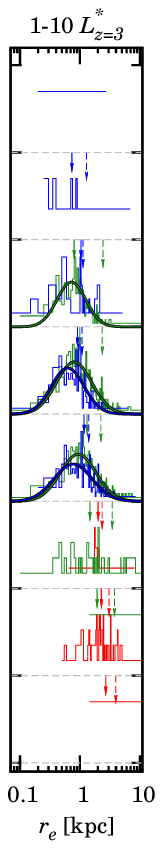}
  \end{center}
  \caption[]{{\footnotesize Distribution of $r_e$ in different $L_{\rm UV}$ bins, $0.12-0.3$ (left), $0.3-1$ (middle), and $1-10$ (right) $L_{\rm UV}/L_{z=3}^*$. Each row displays galaxies from $z=0-1$ (bottom) to $z=8$ (top). The red, green, and blue histograms indicate distribution of $r_e^{\rm Opt}$ and $r_e^{\rm UV}$ for the SFGs, and $r_e^{\rm UV}$ for the LBGs, respectively. The solid curves denote the best-fit log-normal functions for these histograms. The solid and dashed arrows present the median and average values of $r_e$ with the color coding same as the curves. The y-axis is arbitrary. }}
  \label{fig_re_hist}
\end{figure*}

We examine redshift evolution of median, average, and modal $r_e$ of our galaxies to evaluate statistical differences and selection biases. We define four $L_{\rm UV}$ bins for these analyses. The $L_{\rm UV}$-bins are $1-10$, $0.3-1$, $0.12-0.3$, and $0.048-0.12$ $L_{\rm UV}/L_{z=3}^*$, where $L_{z=3}^*$ is the characteristic UV luminosity of LBGs at $z\sim3$ \citep[$M_{\rm UV}=-21$, ][]{1999ApJ...519....1S}\footnote{The $L_{\rm UV}$-bins are the same as in previous studies \citep[e.g., ][]{2010ApJ...709L..21O}. The LBGs in the faintest $L_{\rm UV}$ bin are used only for the stacking analysis (Section \ref{subsec_results_muv_re}).}. To investigate the $r_e$ distribution shape, in Figure \ref{fig_re_hist_large} we plot the $r_e$ distribution of SFGs and LBGs at $z\sim1-6$ in the bin of $L_{\rm UV}=0.3-1\,L_{z=3}^*$ that has good $r_e$ measurement accuracies whose typical reduced $\chi^2$ values are the smallest among the $L_{\rm UV}$ bins of the $r_e$ measurements. We fit the $r_e$ with the log-normal distribution,

\begin{equation}\label{eq_lognormal}
 p(r_e) = \frac{1}{r_e\sigma_{\ln{r_e}}\sqrt{2\pi}} \exp \biggl[ -\frac{\ln^2 (r_e/\overline{r_e})}{2\sigma^2_{\ln{r_e}}} \biggr]
\end{equation}

\noindent where $\overline{r_e}$ and $\sigma_{\ln{r_e}}$ are the peak of $r_e$ and the standard deviation of $\ln{r_e}$, respectively. We fit the log-normal functions to the $r_e$-distribution data with two free parameters of $\overline{r_e}$ and $\sigma_{\ln{r_e}}$, and present the best-fit log-normal functions in Figure \ref{fig_re_hist_large} for the data of good statistics, the SFGs at $z\sim 1-3$ and the LBGs at $z\sim 4-6$ in the $L_{\rm UV}=0.3-1.0 L^*_{\rm UV}$ bin. The $r_e$ distributions of the high-$z$ star-forming galaxies are well represented by the log-normal distribution. The reduced $\chi^2$ values are $0.006$, $0.003$, $0.004$, $0.005$, and $0.011$ for the SFGs at $z=1-2$ and $2-3$, and the LBGs at $z\sim4$, $5$, and $6$, respectively. Figure \ref{fig_re_hist} is the same as Figure \ref{fig_re_hist_large}, but for all of our galaxies. Figure \ref{fig_re_hist} indicates that the $r_e$ distributions are well fitted by the log-normal functions in the wide ranges of redshift, $z\sim 0-6$, and the UV luminosity, $\sim 0.12-10L^*_{\rm UV}$. Note that log-normal functions cannot be fitted to the data of the $z\gtrsim 7$ galaxies and some low-$z$ galaxies in Figure \ref{fig_re_hist}, due to the small statistics. Moreover, the fitting result of $z\sim 0-1$ is only obtained for the $r_e^{\rm Opt}$ distribution in the luminosity bin of $0.12-0.1L^*_{\rm UV}$ because of the poor statistics of the other-luminosity bin data.

\begin{figure*}[t!]
  \begin{center}
    \includegraphics[width=120mm]{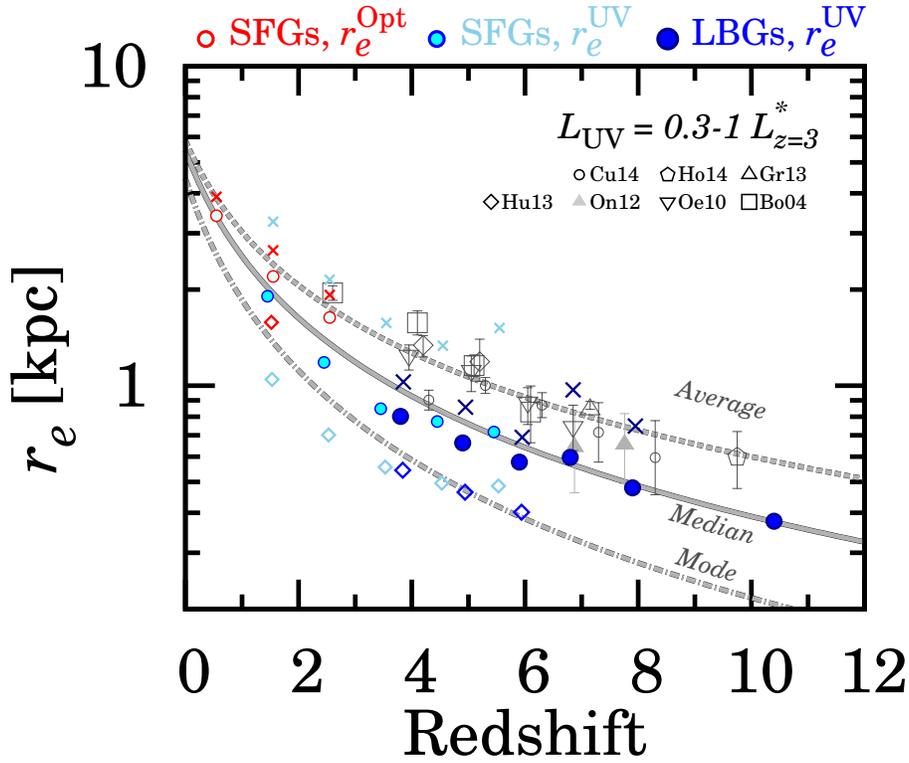}
  \end{center}
  \caption[]{{\footnotesize Difference of the size evolution results based on the average (crosses), median (filled circles), and modal (open diamonds) values of $r_e$ in the bin of $L_{\rm UV} = 0.3-1 L_{z=3}^*$. The red, cyan, and blue filled symbols indicate $r_e^{\rm Opt}$ and $r_e^{\rm UV}$ for the SFGs and $r_e^{\rm UV}$ for the LBGs, respectively. The error bars for our $r_e$ are not plotted for clarity, although these errors are included for estimating $\beta_z$. The solid, dashed, and dot-dashed curves denote the best-fit size evolution for the average, median, and modal $r_e$ values, respectively, in the linear space. Note that these differences of the statistical results are found in the linear space of $r_e$, because the $r_e$-distributions follow the log-normal functions (see the text). The $r_e$ values for LBGs in the literature are plotted with gray symbols (open circles; \citealt{2014arXiv1409.1832C}, open pentagon; \citealt{2014arXiv1406.1180H}, open diamonds; \citealt{2013ApJ...765...68H}, filled triangles; \citealt{2013ApJ...777..155O}, open triangle; \citealt{2012AA...547A..51G}, open inverse-triangles; \citealt{2010ApJ...709L..21O}, and open squares; \citealt{2004ApJ...611L...1B}). }}
  \label{fig_z_re_lbg_comparison}
\end{figure*}

Because the $r_e$-distributions follow the log-normal functions, the average, median, and modal values of $\ln r_e$ should be the same. However, in the previous studies, the size evolution is discussed with the average, median, and modal values of $r_e$ in the linear space \citep[e.g., ][]{2014arXiv1406.1180H,2013ApJ...777..155O,2012AA...547A..51G,2010ApJ...709L..21O,2004ApJ...611L...1B}. Here we obtain $r_e$ measurements with different statistics choices in the linear space, following the previous studies, and evaluate the differences of the size evolution results. We derive size growth rates based on average, median, and modal $r_e$ in a bin of $0.3-1\,L_{\rm UV}/L_{z=3}^*$, estimating  the modal $r_e$ by fitting size distributions with a log-normal function. In Figure \ref{fig_z_re_lbg_comparison}, we compare our $r_e$ measurements with those of the previous studies that apply the different statistics. \footnote{Because there are only three LBGs at $z\sim 10$, the weighted average $r_e$ is only derived for our $z\sim 10$ LBGs. Note that the $z\sim 10$ data is presented in Figure \ref{fig_z_re_lbg_comparison}, but that the data is not used to derive the size evolution function below.} We confirm that our results are consistent with those of the previous studies. Moreover, Figure \ref{fig_z_re_lbg_comparison} indicates that galaxy sizes decrease from $z\sim0$ to $\sim6$ in any statistical choices of average, median, and mode.

%%%%%%%%%%%%%%%%%%%%%%%%%%%%%%%%%%%%%%%%%%%%%%

We fit $r_e = B_z (1+z)^{\beta_z}$ for the average, median, and modal $r_e$ values given by our and previous studies, where $B_z$ and $\beta_z$ are free parameters. The fitting is performed for the combination of $r_e^{\rm UV}$ and $r_e^{\rm Opt}$ as well as for $r_e^{\rm UV}$ only. Table \ref{table_fits} summarizes the best-fit $B_z$ and $\beta_z$ values. Table \ref{table_previous_studies} is a summary of the samples and $\beta_z$ values from our and previous studies for LBGs $z\gtrsim4$. Our average, median, and modal $r_e$ values scale as $\propto (1+z)^{-1\sim-1.3}$, indicating that, again, the choices of statistics in $r_e$ measurements give no significant impacts on size growth rates. This conclusion is consistent with the result that $\sigma_{\ln{r_e}}$ shows no significant evolution as discussed in Section \ref{subsubsec_discuss_LND}.

Most previous studies have employed average values for representative $r_e$. However, Figure \ref{fig_re_hist} indicates that the median measurements trace the typical galaxy sizes parametrized by $\overline{r_e}$ better than the average values. Because the small samples of $z\gtrsim7$ galaxies do not allow us to estimate modal $r_e$ values, we use median values for our main analyses, unless otherwise specified. 

Figure \ref{fig_z_re_lbg_comparison} compares the $r_e$ values of SFGs and LBGs at $z\sim4-6$ with a bright UV luminosity. In any statistics choices, we find that the $r_e$ values of SFGs and LBGs are comparable within the scatters of $\lesssim30$\%. These results indicate that star-forming galaxies selected by photo-$z$ and dropout techniques statistically give the similar $r_e$ values, and that the bias from the different selection techniques is as small as $\lesssim30$\% in the $r_e$ determination.

%% RESULTS
%%%%%%%%%%%%%%%%%%%%%%%%%%%%%%%%%%%%%%%%%%%%%%%%
%%%%%%%%%%%%%%%%%%%%%%%%%%%%%%%%%%%%%%%%%%%%%%%%
\section{RESULTS}\label{sec_results}

%% Sersic Index
%%%%%%%%%%%%%%%%%%%%%%%%%%%%%%%%%%%%%%%%%%%%%%%%
\subsection{S\'ersic Index}\label{subsec_results_z_n}

A S\'ersic index represents the SB profiles of galaxies. A high $n$ means a cuspier SB distribution, indicating the existence of a central bulge. On the other hand, a lower $n$ suggests a disk-like light profile with a flatter SB distribution at the central galactic region. The S\'ersic index depends on observed wavebands and stellar populations (e.g., color), which have been revealed by detailed structural analyses with multiple passbands for local galaxies \citep[e.g., ][]{2013MNRAS.430..330H,2013MNRAS.435..623V,2014MNRAS.444.3603V}. \citet{2014MNRAS.441.1340V} have reported that $n$ tends to be larger in redder bands for blue galaxies due to a bulge component with old stellar ages and/or dust attenuation at the central region.

\begin{figure*}[t!]
  \begin{center}
    \includegraphics[width=230mm, angle=90]{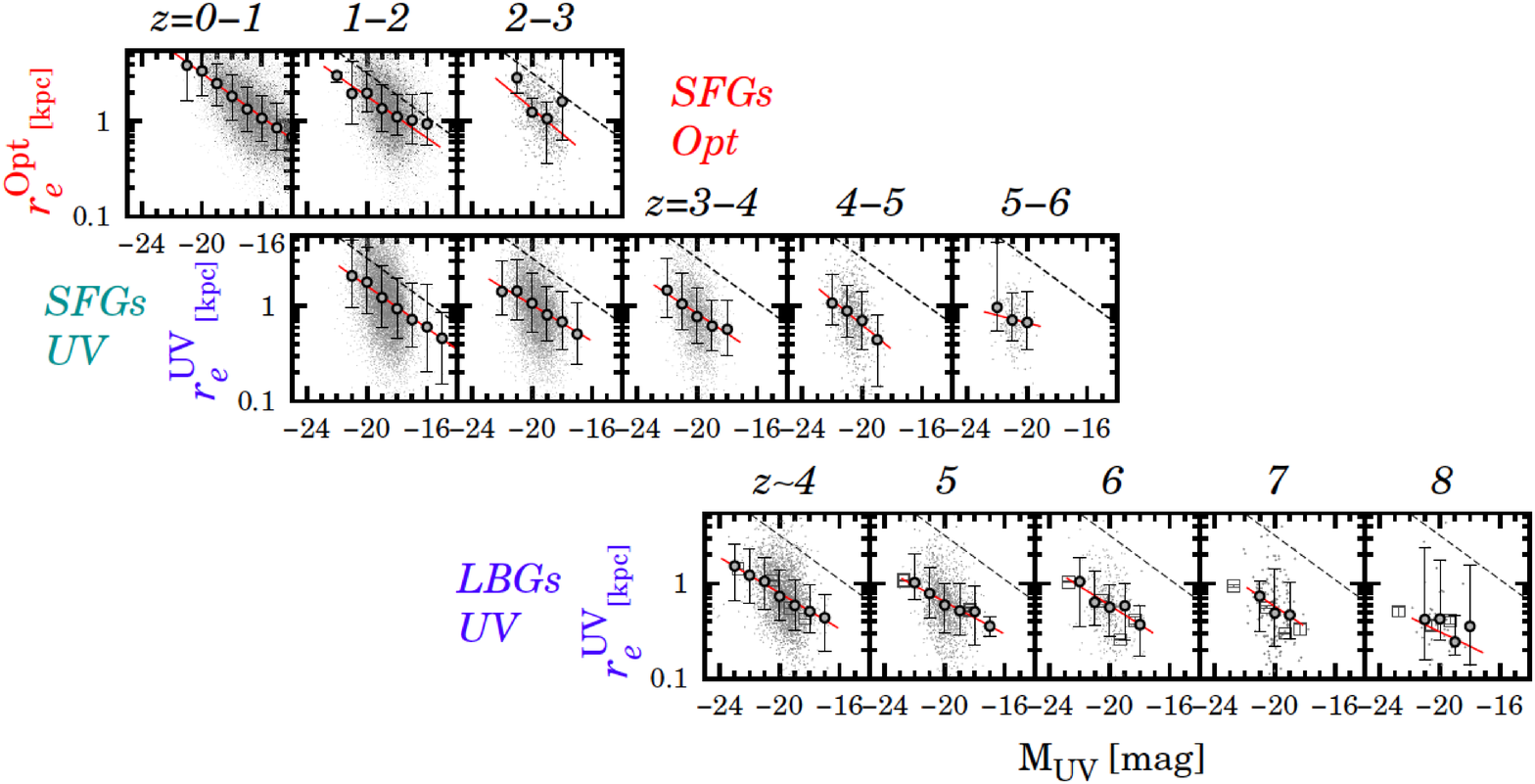}
  \end{center}
  \caption[]{{\footnotesize Effective radius $r_e$ and UV magnitude $M_{\rm UV}$ relation at $z\sim0-8$. The top, middle, and bottom panels represent $r_e^{\rm Opt}$ and $r_e^{\rm UV}$ for the SFGs, and $r_e^{\rm UV}$ for the LBGs, respectively. The redshifts for the relations are labeled at the top of the panels. The red lines denote the best-fit power-law functions of $r_e\propto L_{\rm UV}\,^\alpha$ for the $r_e$-M$_{\rm UV}$ relations. The best-fit power law for the $z\sim0-1$ SFGs are plotted in the all panels (the dashed lines). The open squares in the bottom panels denote $r_e$ values obtained with the stacked images of LBGs  for the purpose of the cosmological SB dimming effect evaluation (see Section \ref{subsec_results_muv_re}). The gray points with error bars indicate the median $r_e$ and the 16th and 84th percentiles of the distribution. }}
  \label{fig_muv_re_lbg}
\end{figure*}

\begin{figure*}[t!]
  \begin{center}
    \includegraphics[width=85mm]{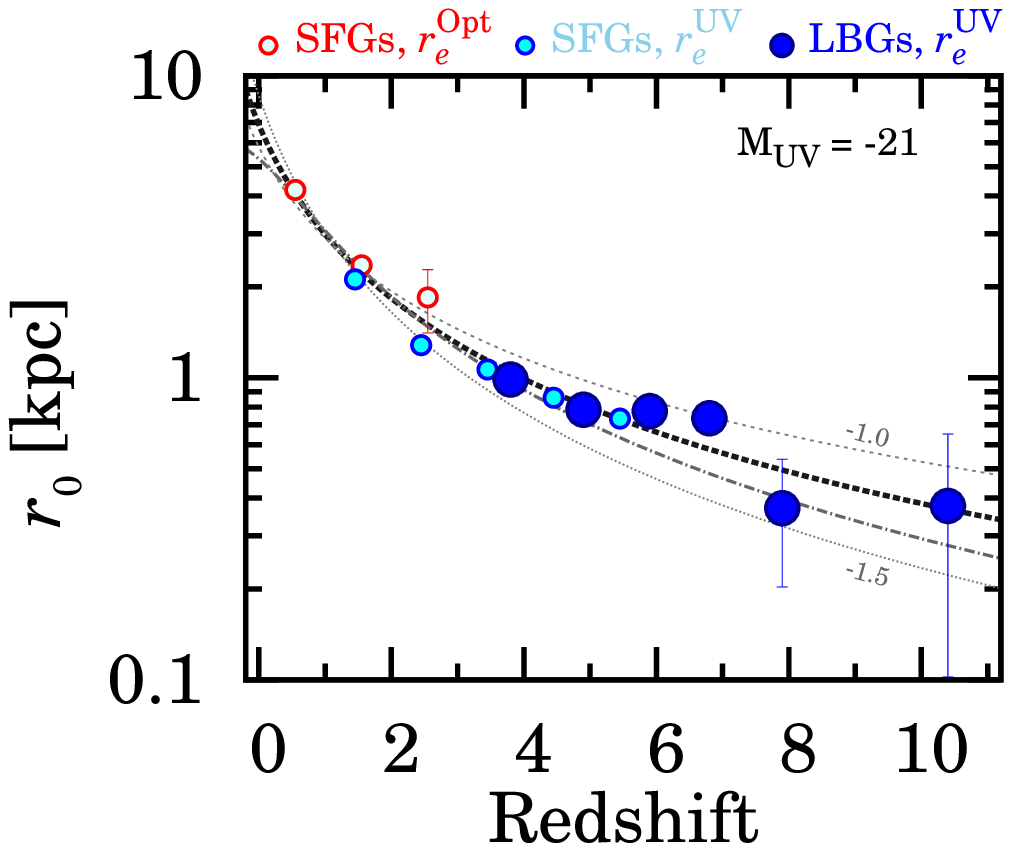}
    \includegraphics[width=85mm]{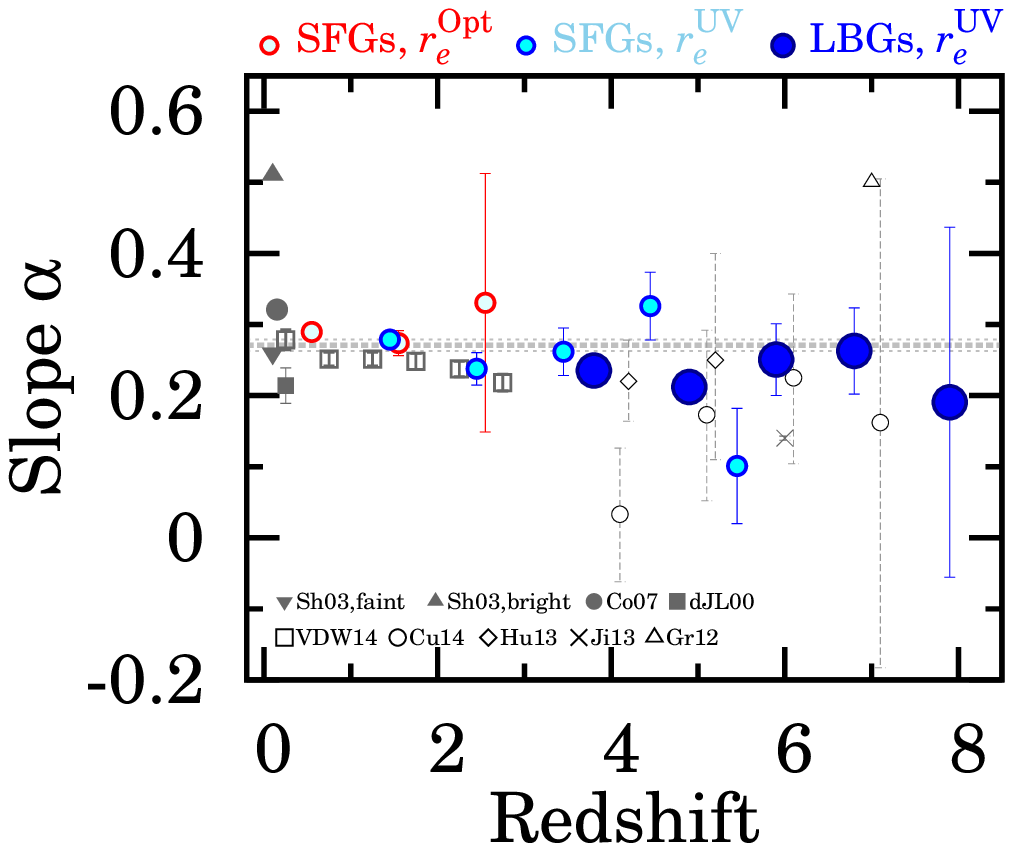}
  \end{center}
  \caption[]{{\footnotesize Results of power-law fits for the $r_e$-M$_{\rm UV}$ relation in Figure \ref{fig_muv_re_lbg}. The red, cyan, and blue filled circles indicate estimates of $r_0$ and $\alpha$ based on $r_e^{\rm Opt}$ and $r_e^{\rm UV}$ for the SFGs, $r_e^{\rm UV}$ for the LBGs, respectively. (Left) Effective radius $r_0$ at $L_{\rm UV} = 1\,L_{z=3}^*$ corresponding to $M_{\rm UV} = -21$. The thin dashed, dotted and thick dashed lines show the best-fit curves with $(1+z)^{-1}$, $(1+z)^{-1.5}$, and $(1+z)^{\beta_z}$, respectively. The dot-dashed line indicates the fit of $r_e\propto H(z)^{\beta_H}$. The best-fit $\beta_z$ and $\beta_H$ values are $-1.20\pm0.04$ and $-0.97\pm0.04$, respectively. (Right) Slope $\alpha$ of $r_e\propto L_{\rm UV}\,^\alpha$ as a function of redshift. The thick dashed and thin gray lines denote the weighted-average value with a $1\sigma$ error, $\alpha=0.27\pm0.01$. The open symbols show $\alpha$ for the SFGs or the LBGs in the literature (open squares assuming Equation \ref{eq_muv_mass_faint}; \citealt{2014ApJ...788...28V}, open circles; \citealt{2014arXiv1409.1832C}, open diamonds; \citealt{2013ApJ...765...68H}, cross; \citealt{2013ApJ...773..153J}, and open triangle; \citealt{2012AA...547A..51G}). The gray filled symbols represent the results for local spiral and/or disk galaxies (filled triangle and inverse-triangle; $n<2.5$ galaxies with $r$-band magnitudes of $M_r\leq -20.91$ and $M_r\geq -20.91$, respectively, in \citealt{2003MNRAS.343..978S}, filled circle; \citealt{2007ApJ...671..203C}, and filled square; \citealt{2000ApJ...545..781D}). }}
  \label{fig_z_slope}
\end{figure*}

\begin{figure*}[t!]
  \begin{center}
    \includegraphics[width=170mm]{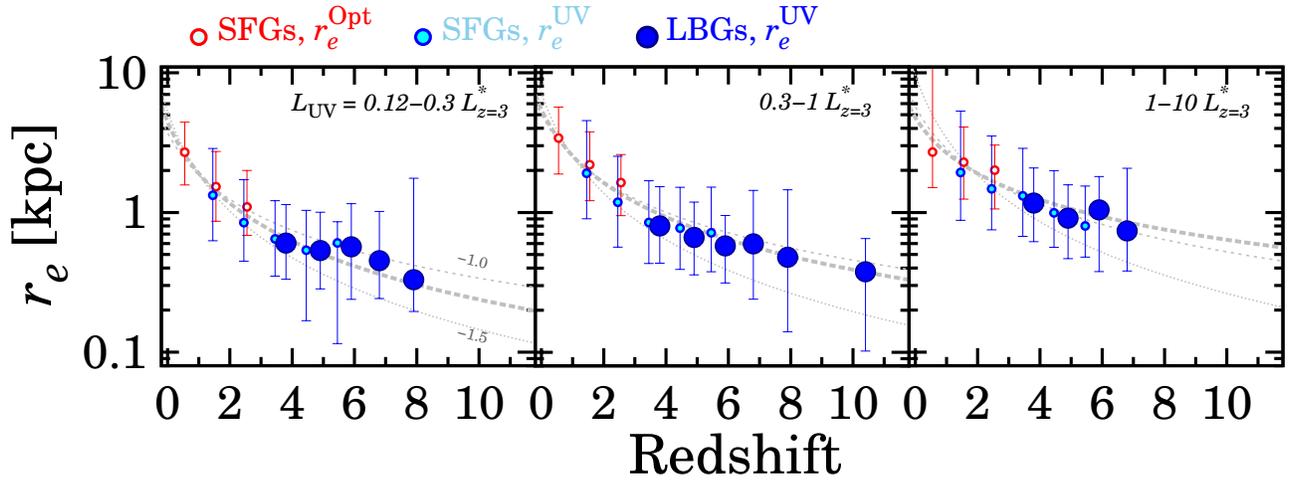}
  \end{center}
  \caption[]{{\footnotesize Redshift evolution of median $r_e$ in different $L_{\rm UV}$ bins, $0.12-0.3$ (left), $0.3-1$ (center), and $1-10$ (right) $L_{\rm UV}/L_{z=3}^*$. The definitions of the symbols and lines are the same as those in Figure \ref{fig_z_slope}. The data points are slightly shifted along $x$-axis for clarity. }}
  \label{fig_z_re_lbg}
\end{figure*}

\begin{figure}[t!]
  \begin{center}
    \includegraphics[width=85mm]{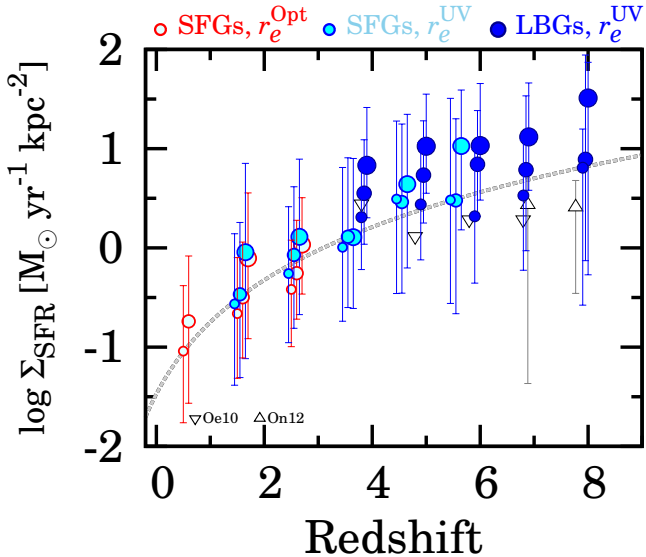}
  \end{center}
  \caption[]{{\footnotesize SFR SD $\Sigma_{\rm SFR}$ as a function of redshift. The definitions of the symbols are the same as those in Figure \ref{fig_z_slope}. The large, middle, and small circles denote $\Sigma_{\rm SFR}$ in the $L_{\rm UV}$ bins of $1-10$, $0.3-1$, and $0.12-0.3$ $L_{\rm UV}/L_{z=3}^*$, respectively. The filled circles are the same as in Figure \ref{fig_z_slope}. The SFR for the LBGs is corrected for dust extinction with two relations of $M_{\rm UV}$-$\beta$ \citep{2014ApJ...793..115B} and IRX-$\beta$ \citep{1999ApJ...521...64M}. The dashed gray lines represent the $\Sigma_{\rm SFG}$ evolution calculated with an SFR of $10\,M_\odot/$yr and the best-fit $r_0$ curve in Figure \ref{fig_z_slope}. The open symbols are taken from the literature on LBGs (triangles; \citealt{2013ApJ...777..155O}, inverse-triangles; \citealt{2010ApJ...709L..21O}). The error bars denote the 16th and 84th percentiles of distribution. The data points are slightly shifted along $x$-axis for clarity. }}
  \label{fig_z_sfrsd_lbg}
\end{figure}

\begin{figure*}[t!]
  \begin{center}
    \includegraphics[width=150mm]{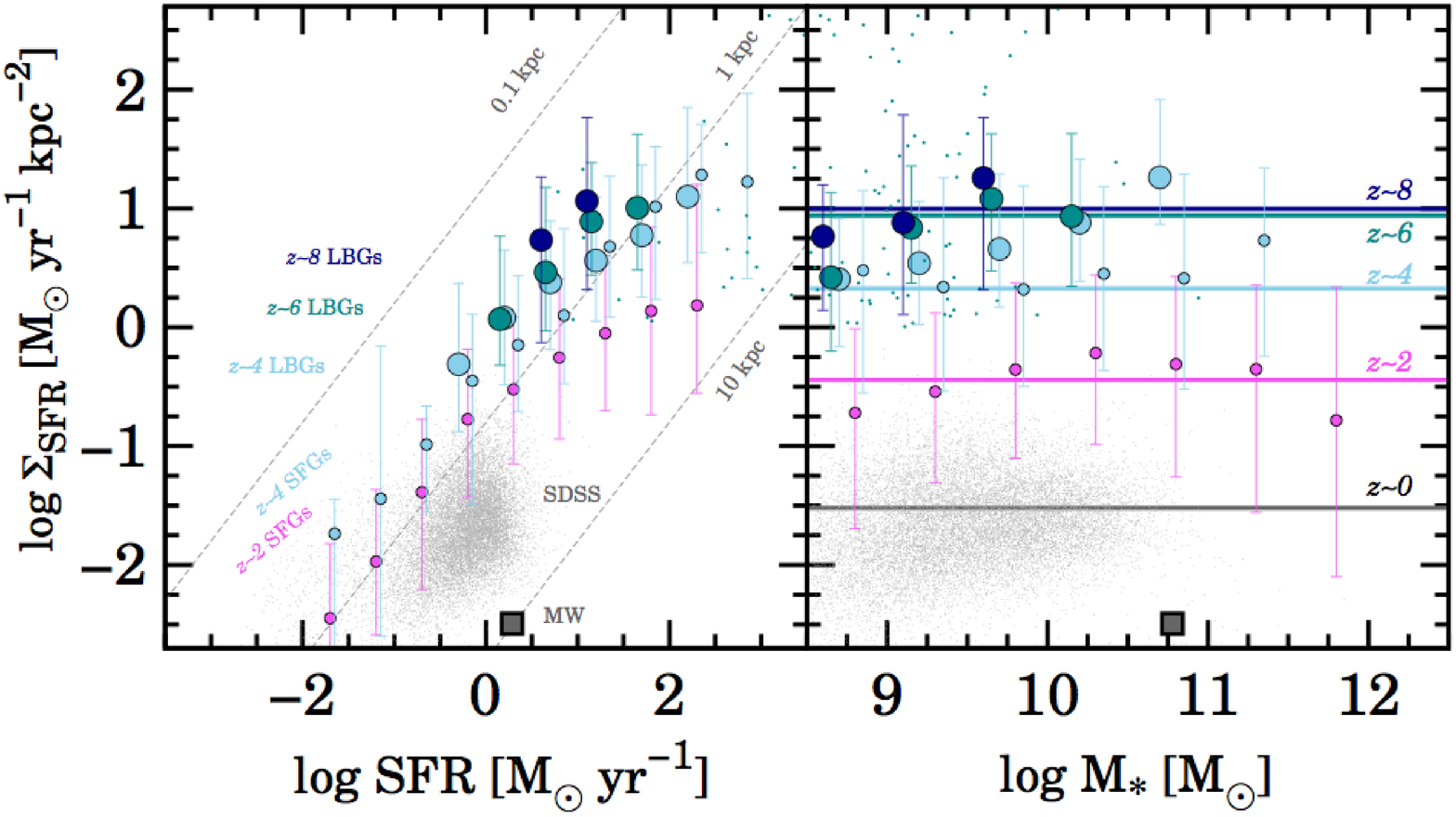}
  \end{center}
  \caption[]{{\footnotesize SFR SD $\Sigma_{\rm SFR}$ as functions of SFR (left) and stellar mass (right). The small magenta and cyan circles indicate median $\Sigma_{\rm SFR}$ values at a given SFR or M$_*$ for the SFGs at $z\sim2$ and $z\sim4$, respectively, based on $r_e^{\rm UV}$. The large circles represent the LBGs at $z\sim4$ (cyan), $z\sim6$ (green), and $z\sim8$ (dark-blue). The dark-blue points denote individual LBGs at $z=7-8$. The SFR for the LBGs is corrected for dust extinction with two relations of $M_{\rm UV}$-$\beta$ \citep{2014ApJ...793..115B} and IRX-$\beta$ \citep{1999ApJ...521...64M}.  The square represents the Milky Way \citep{2012ARA&A..50..531K}. The gray dots indicate SDSS galaxies with an exponential SB distribution from a catalog of \cite{2012MNRAS.421.2277L} whose $\Sigma_{\rm SFR}$ calculated from the SFR and $r_e$ values based on $u$-band magnitudes and single S\'ersic profile fits, respectively. The stellar mass of the SDSS galaxies is taken from \citet{2003MNRAS.341...33K,2004MNRAS.351.1151B,2007ApJS..173..267S}. The dashed lines correspond to constant effective radii of $r_e=0.1, 1, 10$ kpc, from top to bottom. The horizontal lines are the weighted average values of $\Sigma_{\rm SFR}$ in each redshift bin. The error bars denote the 16th and 84th percentiles of the distribution. }}
  \label{fig_sfr_mass_sfrsd}
\end{figure*}

Our results confirm that $n$ values of QGs are significantly higher than those of SFGs at $z\lesssim 2$ in the second-top right panel of Figure \ref{fig_sfg_tile}. For massive SFGs with $\log M_*=10-11$ M$_\odot$, $n$ values monotonically increase from $n\sim1-1.5$ at $z\sim 1$ to $n\sim2-3$ at $z\sim 0$. The evolutionary trend of $n$ for the massive SFGs is similar to that of the QGs at $z\sim0-2$, which is consistent with previous results (see the discussions in \citealt{2013A&A...553A..80P,2009ApJ...699L.178N,2010ApJ...709.1018V}). 

At $2\lesssim z\lesssim3$, $n$ values of the SFGs at the rest-frame optical wavelengths are smaller than those at the rest-frame UV wavelengths slightly by $\Delta n \lesssim 0.5$, which is similar to the results of \citet{2014MNRAS.441.1340V} for local objects.

Interestingly, in Figure \ref{fig_sfg_tile}, we find that typical SFGs have a value of $n\sim1-1.5$ at the wide redshift range of $z\sim 1-6$, albeit with the large scatter of individual galaxies. There is a similar claim made by e.g. \citet{2014ApJ...785...18M}, but only for $z\sim 1-3$ star-forming galaxies (Figure \ref{fig_sfg_tile}). Our results newly suggest that the typical S\'ersic indices of star-forming galaxies are $n\sim1-1.5$ at $z\sim 3-6$.

This constant $n$ guarantees that we use a fixed $n$ value of $1.5$ in the size measurements for LBGs (Section \ref{sec_analysis_size}).

%% Luminosity-Size Relation
%%%%%%%%%%%%%%%%%%%%%%%%%%%%%%%%%%%%%%%%%%%%%%%%
\subsection{Size-Luminosity Relation}\label{subsec_results_muv_re}

We investigate the size-luminosity $r_e$-$L_{\rm UV}$ relation and its dependence on redshift. Figure \ref{fig_muv_re_lbg} and Table \ref{table_muv_re} represent the size-luminosity relation at $z=0-8$ for the SFGs and LBGs, where $L_{\rm UV}$ is presented with $M_{\rm UV}$. We cannot examine the size-luminosity relation at $z\sim 10$, because the number of $z\sim 10$ LBGs is only three. A large area of $\sim910$ arcmin$^2$ in the {\it HST} fields allows us to derive the $r_e$-$L_{\rm UV}$ relation in a wide range of magnitude, $-23 \lesssim M_{\rm UV}\lesssim -17$ mag even for $z\sim4$ LBGs. Figure \ref{fig_muv_re_lbg} shows that $r_e$ has a negative correlation with M$_{\rm UV}$ at $0\lesssim z\lesssim8$. 

The $r_e$-$L_{\rm UV}$ relation is fitted by
 
\begin{equation}\label{eq_re_luv}
r_e = r_0 \Biggl( \frac{L_{\rm UV}}{L_0} \Biggr) ^\alpha, 
\end{equation} 

\noindent where $r_0$ and $\alpha$ are free parameters. The $r_0$ value represents the effective radius at a luminosity of $L_0$, which is similar to the parameter $\gamma$ used in e.g., \citet[][]{2012ApJ...746..162N}. The $\alpha$ value is the slope of the $r_e$-$L_{\rm UV}$ relation. We select $L_0$ to the best-fit Schechter parameter $M^*$ at $z\sim3$ that corresponds to $M_{\rm UV}=-21.0$, following the arguments of \citet{2013ApJ...765...68H}. 

The left panel of Figure \ref{fig_z_slope} shows the redshift evolution of $r_0$ and $\alpha$. We parametrize the size growth rate by fitting $r_0$ with a function of $B_z (1+z)^{\beta_z}$. The best-fit function is $6.9\,(1+z)^{-1.20\pm-0.04}$ kpc, which do not significantly change even with and without the $r_e^{\rm Opt}$ results. We also carry out fitting with a function of $B_H h(z)^{\beta_H}$, where $B_H$ and ${\beta_H}$ are free parameters and $h(z) \equiv H/H_0 = \sqrt{\Omega_m (1+z)^3 + \Omega_\Lambda}$.  Here the fitting of the $h(z)$-form functions are conducted, because these $h(z)$-form functions could be a more realistic physical treatment as claimed by \citet[e.g., ][]{2014ApJ...788...28V}. The fitting results yield the best-fit function of $5.3\,h(z)^{-0.97\pm0.04}$ kpc that is plotted in the left panel of Figure \ref{fig_z_slope}. Although we do not use the $r_e$ estimate of $z\sim 10$ for the fitting, the $z\sim 10$ data point is placed on the the extrapolation of the best-fit function.

\begin{figure*}[t!]
  \begin{center}
    \includegraphics[width=175mm]{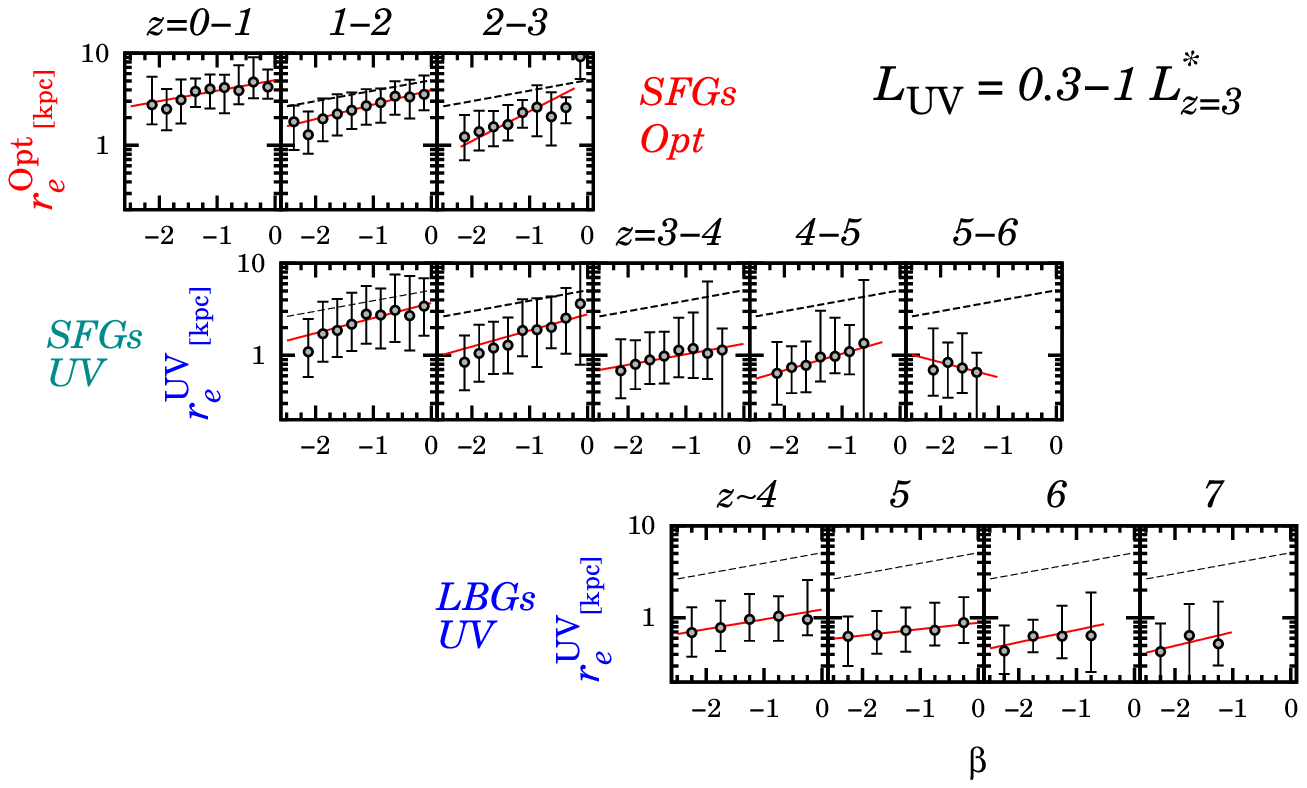}
  \end{center}
  \caption[]{{\footnotesize Relation between effective radius $r_e$ and UV slope $\beta$ for the SFGs and the LBGs with $L_{\rm UV}=0.3-1\,L_{z=3}^*$. The symbols are the same as in Figure \ref{fig_muv_re_lbg}. The red lines denote the best-fit power-law functions of $r_e\propto\beta^c$, where c is a free parameter. The $z\sim8$ relation is not shown here due to the poor statistics. }}
  \label{fig_beta_re}
\end{figure*}

\begin{figure}[t!]
  \begin{center}
    \includegraphics[width=89mm]{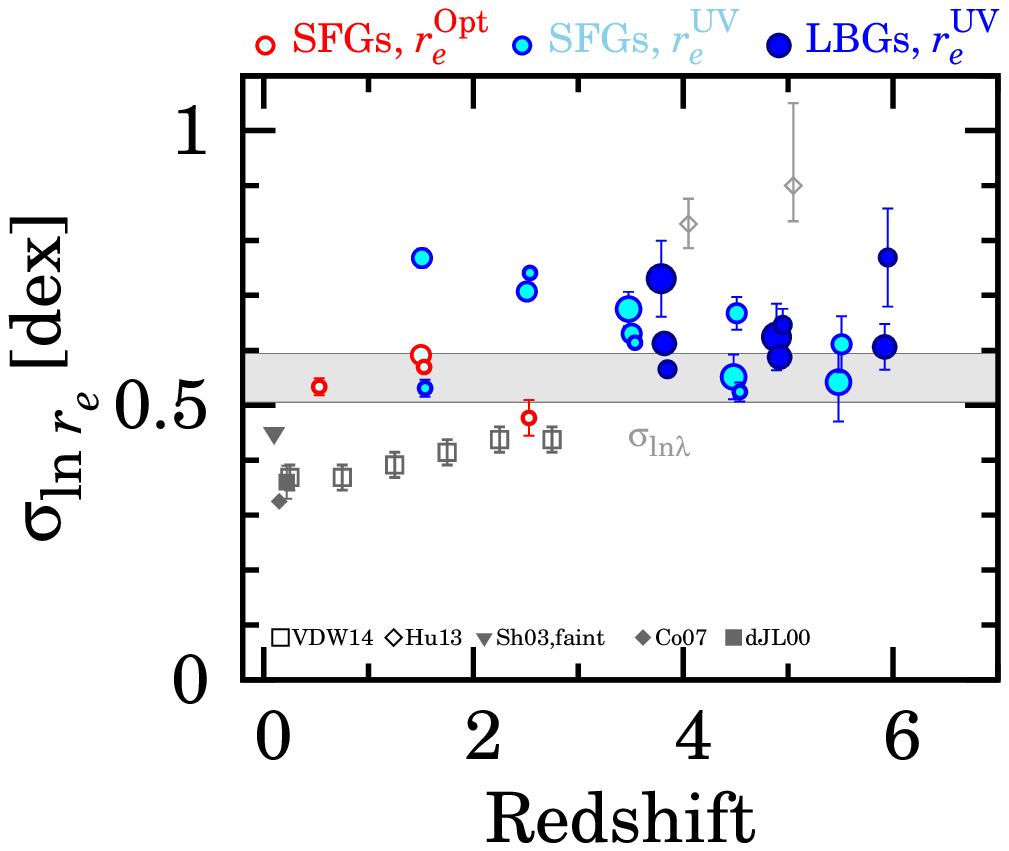}
  \end{center}
  \caption[]{{\footnotesize Standard deviation $\sigma_{\ln{r_e}}$ of the log-normal $r_e$ distribution (Equation \ref{eq_lognormal}) as a function of redshift. The colored symbols are the same as in Figure \ref{fig_z_slope}, but the large, medium-large, and small circles denote $\sigma_{\ln{r_e}}$ in the $L_{\rm UV}$ bins of $1-10$, $0.3-1$, and $0.12-0.3$ $L_{\rm UV}/L_{z=3}^*$, respectively. The shaded region indicates the width of the $\lambda$ distribution, $\sigma_{\ln{\lambda}}$, predicted by N-body simulations \citep[e.g., ][]{2001ApJ...555..240B}. The $\sigma_{\ln{r_e}}$ value at $z\gtrsim7$ is not plotted due to the poor statistics. The gray open symbols show $\sigma_{\ln{r_e}}$ for SFGs or LBGs in the literature (open squares; \citealt{2014ApJ...788...28V}, open diamonds; \citealt{2013ApJ...765...68H}). The gray filled symbols represent the results for local spiral or disk galaxies (filled inverse-triangle for $n<2.5$ galaxies with $M_r\geq -20.91$; \citealt{2003MNRAS.343..978S}, filled circle; \citealt{2007ApJ...671..203C}, filled square; \citealt{2000ApJ...545..781D}). }}
  \label{fig_z_width}
\end{figure}

The evolution of $r_0$ is similar to those of the median $r_e$ values that are presented in Figure \ref{fig_z_re_lbg_comparison}. Here we plot $r_e$ as a function of redshift in Figure \ref{fig_z_re_lbg}, which is the same as Figure \ref{fig_z_re_lbg_comparison}, but for the median $r_e$ values of three different UV luminosity samples. We fit the functions and find that the best-fit $\beta_z$ are $-1.22\pm0.05$, $-1.10\pm0.06$, and $-0.84\pm0.11$ in $L_{\rm UV}/L_{z=3}^*=0.12-0.3, 0.3-1$, and $1-10$, respectively (Table \ref{table_fits}). The best-fit $\beta_z$ values are comparable to the one of $r_0$.  

In contrast to the $r_e$ evolution, there is no significant evolution of $\alpha$ (eq. \ref{eq_re_luv}) at $z=0-8$ found in the right panel of Figure \ref{fig_z_slope}. We calculate the weighted-average value of $\alpha$ with our data points over $z=0-8$, and obtain $\alpha=0.27\pm 0.01$. Figure \ref{fig_z_slope} compares the $\alpha$ estimates of $z=0-8$ obtained in the previous studies. The $\alpha$ measurements of local spiral and/or disk galaxies are comparable to $\alpha \sim 0.27$ \citep{2003MNRAS.343..978S,2007ApJ...671..203C,2000ApJ...545..781D}. At $z=0-3$, \citet{2014ApJ...788...28V} have revealed that the slopes of size-stellar mass relation do not evolve. Adopting eq. (\ref{eq_muv_mass_faint}) to calculate $L_{\rm UV}$ from the stellar masses, we obtain the $r_e$-$L_{\rm UV}$ relation and evolution similar to our results. At $z>4$, there are several $\alpha$ measurements reported by \citet{2014arXiv1409.1832C,2013ApJ...765...68H,2013ApJ...773..153J,2012AA...547A..51G}. However, these data points of $\alpha$ are largely scattered (the right panel of Figure \ref{fig_z_slope}). Nevertheless, our $\alpha$ values fall within the scatter of the previous measurements.

Our results of the $r_e$ (or $r_0$) evolution and the constant $\alpha$ suggest that the $r_e$-$L_{\rm UV}$ relation of star-forming galaxies is unchanged but with a decreasing offset of $r_e$ from $z=0$ to $8$. Because the morphological evolution trend of star-forming galaxies is simple, our results benefit to studies using Monte-Carlo simulations for luminosity function determinations that require an assumption of high-$z$ galaxy sizes \citep[e.g., ][]{2014arXiv1408.6903I,2014arXiv1409.1228O}. Moreover, these morphological evolution trends are important constraints on parameters of galaxy formation models.

Note that there is a possible source of systematics given by the cosmological SB dimming effect by which we would underestimate $r_e$ (Section \ref{sec_analysis_size}). To estimate the effect of the cosmological SB dimming, we measure $r_e$ of $z\sim4-8$ LBGs with stacked images that accomplish the detection limit deeper than the individual images by a factor of $\sim20-30$. The $r_e$ values measured in the stacked images roughly reproduce the size-luminosity relation of Figure \ref{fig_muv_re_lbg}, suggesting that there are no signatures of systematics in the $r_e$ values measured by our {\tt GALFIT} profile fitting technique. There is another possibility of the cosmological SB dimming effect. If there exist a large population of diffuse high-$z$ galaxies that are not identified in our {\it HST} images, we would underestimate the $r_e$ values. However, it is unlikely that such a diffuse high-$z$ population exists. This is because the luminosity functions of $z\sim 4-6$ LBGs derived with {\it HST} data agree with those obtained by ground-based observations \citep{2006AJ....132.1729B} whose PSF's FWHM is $\sim1^{\prime\prime}$ corresponding to $3-4$ kpc in radius at $z\sim 4-6$. In other words, at these redshifts, there is no diffuse population with a radius up to $\sim 3-4$ kpc that is significantly larger than our size measurements of $r_e\lesssim 1$ kpc (see, e.g., Figure \ref{fig_re_hist_large}). We therefore conclude that our results of size measurements are not significantly changed by the cosmological SB dimming effect.

%% SFR Surface Density
%%%%%%%%%%%%%%%%%%%%%%%%%%%%%%%%%%%%%%%%%%%%%%%%
\subsection{SFR Surface Density}\label{subsec_results_z_sfrsd}

We examine the redshift evolution of SFR SD $\Sigma_{\rm SFR}$. Figure \ref{fig_z_sfrsd_lbg} shows $\Sigma_{\rm SFR}$ as a function of redshift. Figure \ref{fig_z_sfrsd_lbg} is the same as Figure \ref{fig_sfg_tile}, but for all of our galaxies up to $z=8$ with the binning of $L_{\rm UV}$ values. Figure \ref{fig_z_sfrsd_lbg} shows that $\Sigma_{\rm SFR}$ gradually increases by redshift from $z\sim 0$ to $8$. This evolutional trend and the $\Sigma_{\rm SFR}$ values are consistent with those of $z\sim4-8$ previously reported by \cite[e.g., ][]{2010ApJ...709L..21O,2013ApJ...777..155O}.  Our results of the $\Sigma_{\rm SFR}$ evolution suggests that $\Sigma_{\rm SFR}$ of typical high-$z$ galaxies continuously increases from $z\sim 0$ to $8$.

In Figure \ref{fig_z_sfrsd_lbg}, we also find that the increase rate per redshift becomes small at $z\gtrsim4$ in the regime of $\log{\Sigma_{\rm SFR}}\sim0.5-1$ M$_\odot$ yr$^{-1}$ kpc$^{-2}$. We obtain the $\Sigma_{\rm SFR}$ evolution curve using the eq. (\ref{eq_sfrsd}) with the inputs of the best-fit function $r_0=6.9\,(1+z)^{-1.20\pm-0.04}$ (Section \ref{subsec_results_muv_re}) and the SFR estimated from the $L_{\rm UV}$ value via equation (\ref{eq_sfr}). Figure \ref{fig_z_sfrsd_lbg} presents the $\Sigma_{\rm SFR}$ evolution curve. As expected, the $\Sigma_{\rm SFR}$ evolution curve follows the $\Sigma_{\rm SFR}$ data points. In other words, the slow $\Sigma_{\rm SFR}$ evolution at $z\gtrsim4$ is explained by the simple power-law galaxy size evolution of $r_0=6.9\,(1+z)^{-1.20\pm-0.04}$.

In Figure \ref{fig_sfr_mass_sfrsd}, we examine the dependence of $\Sigma_{\rm SFR}$ on SFR and $M_*$. The left and right panels of Figure \ref{fig_sfr_mass_sfrsd} show $\Sigma_{\rm SFR}$ as functions of SFR and $M_*$, respectively. For comparison, we also plot SDSS galaxies with an exponential SB profile in \citet{2012MNRAS.421.2277L} and the Milky-Way \citep{2012ARA&A..50..531K}. These local galaxies are placed in the regime of low $\Sigma_{\rm SFR}$ values. Obviously, Figure \ref{fig_sfr_mass_sfrsd} reproduces the result of Figure \ref{fig_z_sfrsd_lbg} that $\Sigma_{\rm SFR}$ is typically higher for high-$z$ galaxies than low-$z$ galaxies. In the $\Sigma_{\rm SFR}$-SFR diagram of Figure \ref{fig_sfr_mass_sfrsd}, $\Sigma_{\rm SFR}$ positively correlates with SFR. This is because the $\Sigma_{\rm SFR}$ and SFR values are related by eq. (\ref{eq_sfrsd}). The slopes of $\Sigma_{\rm SFR}$-SFR relation appear similar at $z\sim 2-8$. On the other hand, we find that the $\Sigma_{\rm SFR}$-$M_*$ diagram of Figure \ref{fig_sfr_mass_sfrsd} shows no strong dependence of $\Sigma_{\rm SFR}$ on $M_*$ \citep[see also, e.g. ][]{2011ApJ...742...96W}. These two diagrams suggest that $\Sigma_{\rm SFR}$ increases towards high-$z$, keeping the similar $\Sigma_{\rm SFR}$-SFR and $\Sigma_{\rm SFR}$-$M_*$ relations over $z\sim 2-8$.

%% UV slope beta - size relation 
%%%%%%%%%%%%%%%%%%%%%%%%%%%%%%%%%%%%%%%%%%%%%%%%
\subsection{Size-UV Slope $\beta$ Relation}\label{subsec_beta_re}

We derive the $r_e$-UV slope $\beta$ relation to investigate the dependence of galaxy sizes on stellar population. The $\beta$ parameter is defined by $f_\lambda\propto\lambda^\beta$ where $f_\lambda$ is a galaxy spectrum at $\sim 1500-3000$\AA, which is a coarse indicator of the stellar population and extinction of galaxies. 
A small $\beta$ means a blue spectral shape, suggesting young stellar ages, low metallicity, and/or dust extinction. 

For the SFGs, we calculate $\beta$ via 
 
\begin{equation}
\beta = - \frac{m_{1700} - m_{2800}}{2.5\log{1700/2800}} - 2, 
\end{equation}

\noindent where $m_{1700}$ and $m_{2800}$ are the total magnitudes at wavelengths of $1700$ and $2800$\,\AA\, in the rest-frame, respectively. These magnitudes are taken from the catalogue of \citet{2014arXiv1403.3689S}. For the $z\sim4$, $5$, and $6$ LBGs, we derive $\beta$, fitting the function of $f_\lambda\propto\lambda^\beta$ to the magnitude sets of $i_{775}I_{814}z_{850}Y_{105}J_{125}$, $z_{850}Y_{105}J_{125}H_{160}$, and $Y_{105}J_{125}H_{160}$, respectively, in the same manner as \cite{2014ApJ...793..115B}.  For the $z\sim7$ and $8$ LBGs, we estimate $\beta$ using

\begin{eqnarray}\label{eq_beta}
&\beta& = -2.0 + 4.59(J_{125} - H_{160})\, ({\rm for}\,z\sim7), \\
&\beta& = -2.0 + 8.68(JH_{140} - H_{160})\, ({\rm for}\,z\sim8). 
\end{eqnarray}

Figure \ref{fig_beta_re} represents the $r_e$-$\beta$ relation in the bin of $L_{\rm UV}/L_{z=3}^*=0.3-1$. We find that $L_{\rm UV}$-$beta$ relation is poorly determined for the $z\sim8$ LBGs, due to the small statistics, and the $z\sim 8$ result is not presented. In Figure \ref{fig_beta_re}, we identify clear trends that smaller galaxies have a bluer UV spectral shape at $0\lesssim z\lesssim7$. This is consistent with the results of $z\sim 6-8$ LBGs reported by \citet{2014arXiv1410.1535K}. This $r_e$-$\beta$ correlation indicates that young and forming galaxies have typically a small size. We find a negative correlation between $r_e$ and $\beta$ for the $z=5-6$ SFGs. The negative-correlation trend appears simply due to the small sample, which is not statistically significant.

%% DISCUSSION
%%%%%%%%%%%%%%%%%%%%%%%%%%%%%%%%%%%%%%%%%%%%%%%%
%%%%%%%%%%%%%%%%%%%%%%%%%%%%%%%%%%%%%%%%%%%%%%%%
\section{DISCUSSION}\label{sec_discussion}

%% 
%%%%%%%%%%%%%%%%%%%%%%%%%%%%%%%%%%%%%%%%%%%%%%%%
\subsection{The $r_e$ Distribution and SHSR:\\ 
Implications for Host DM Halos and Disks}\label{subsec_discuss_main}

Here we investigate the properties of the $r_e$ distributions in Section \ref{subsubsec_discuss_LND}, and estimate SHSRs in Section \ref{subsubsec_discuss_SHSR}. Combining these results and theoretical models, we discuss the host DM halos and the stellar dynamics in Section \ref{subsubsec_discuss_disk}.

%%%%%%%%%%%%%%%%%%%%%%%%%%%%%%%%%%%%%%%%%%%%%%%%
\subsubsection{Log-Normal Distribution of $r_e$}\label{subsubsec_discuss_LND}

In Section \ref{subsec_bias_statistics}, we find that the $r_e$ distributions of our galaxies are well fitted by the log-normal functions in the wide-range of redshift, $z\sim 0-6$, and luminosity.

Figure \ref{fig_z_width} shows the best-fit $\sigma_{\ln{r_e}}$ values as a function of redshift. Size measurement uncertainties $\sigma_{\ln{r_e}, {\rm err}}$ would broaden the width of the $r_e$-distribution. We estimate typical $\sigma_{\ln{r_e}, {\rm err}}$ in each $z$ and $L_{\rm UV}$ bin. We correct $\sigma_{\ln{r_e}}$ for the size measurement uncertainties through $\sigma_{\ln{r_e}} = (\sigma_{\ln{r_e}, {\rm obs}}\,^2 - \sigma_{\ln{r_e}, {\rm err}}\,^2)^{0.5}$, where $\sigma_{\ln{r_e}, {\rm obs}}$ is the observed width of the $r_e$ distribution. We find that $\sigma_{\ln{r_e}}$ values fall in the range of $\sim 0.45-0.75$ with no clear evolutional trend at $z\sim 0-6$. Our $\sigma_{\ln{r_e}}$ values are slightly larger than the estimates for local disks in \citet{2003MNRAS.343..978S,2000ApJ...545..781D,2007ApJ...671..203C} and for late-type galaxies at $z\sim0-3$ in \citet{2014ApJ...788...28V}. These differences would be explained by the choices of the wavelengths for the galaxy size measurements, because these previous studies measure galaxy sizes in the rest-frame optical wavelength. In fact, if we change from the rest-frame UV-luminosity to optical wavelength sizes for the size distribution, we obtain moderately small $\sigma_{\ln{r_e}}$ values. However, there still remain the differences of $\sim 20-30$\% beyond the error bars in Figure \ref{fig_z_width}. These $\sim 20-30$\% differences are probably explained by the sample and measurement technique differences. We also compare the $\sigma_{\ln{r_e}}$ estimates of $z\sim4-5$ LBGs given by \citet{2013ApJ...765...68H}, and find a moderately large difference by a factor of 1.5. However, the scatters of our measurements and the statistical uncertainties of \citealt{2013ApJ...765...68H}'s estimates are too large to conclude the differences.

%%%%%%%%%%%%%%%%%%%%%%%%%%%%%%%%%%%%%%%%%%%%%%%%
\subsubsection{SHSR} \label{subsubsec_discuss_SHSR}

We estimate the SHSRs that are defined with the ratio of $r_e / r_{\rm vir}$, where $r_{\rm vir}$ the virial radius of a host DM halo. 

\begin{figure}[t!]
  \begin{center}
    \includegraphics[width=83mm]{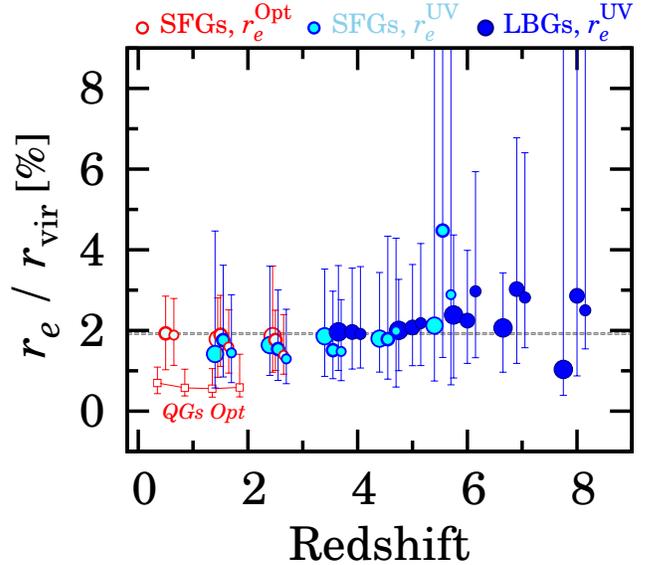}
  \end{center}
  \caption[]{{\footnotesize Median SHSR, $r_e/r_{\rm vir}$, as a function of redshift. The colored symbols are the same as in Figure \ref{fig_z_slope}, but the large, medium-large, and small circles denote $r_e/r_{\rm vir}$ values in the $L_{\rm UV}$ bins of $1-10$, $0.3-1$, and $0.12-0.3$ $L_{\rm UV}/L_{z=3}^*$, respectively. The horizontal dashed line indicates a weighted mean of $\left< r_e/r_{\rm vir} \right>$ in the $0.3-1\,L_{\rm UV}/L_{z=3}^*$ bin. The red squares denote the QGs with $\log{M_*}=10.5-11\, M_\odot$. The virial mass of host DM halos is derived from the results of \citet{2013ApJ...770...57B}. }}
  \label{fig_z_rerdm_lbg}
\end{figure}

The $r_{\rm vir}$ value is calculated by 

\begin{equation}\label{eq_virial}
  r_{\rm vir} = \Biggl(  \frac{2GM_{\rm vir}}{\Delta_{\rm vir}\Omega_{\rm m}(z)H(z)^2} \Biggr)^{1/3}, 
\end{equation}

\noindent where $\Delta_{\rm vir}=18\pi^2+82x-39x^2$ and $x=\Omega_m (z) - 1$ \citep{1998ApJ...495...80B}. We obtain the virial mass of a DM halo, $M_{\rm vir}$, from stellar mass, $M_*$, of individual galaxies by using the relation determined by the abundance matching analyses \citep{2010ApJ...717..379B,2013ApJ...770...57B}. Figure \ref{fig_z_rerdm_lbg} shows $r_e/r_{\rm vir}$ as a function of redshift and its dependence on $L_{\rm UV}$ at $z\sim0-8$. The $z\sim10$ data point is omitted due to small statistics. In Figure \ref{fig_z_rerdm_lbg}, we find that $r_e/r_{\rm vir}$ is $\sim 2$\% for the star-forming galaxies and $\sim 0.5$\% for the QGs. Interestingly, $r_e/r_{\rm vir}$ of the star-forming galaxies is almost constant with redshift, albeit with the large uncertainties at $z\gtrsim5$. The no significant evolution of $r_e/r_{\rm vir}$ is reported by \citet{2014arXiv1410.1535K} based on a compilation of data from the literature for star-forming galaxies at $z\gtrsim 2$. Our systematic structural analyses confirm the report of no large evolution seamlessly from $z\sim 0$ with the homogenous data sets and the same analysis technique over the wide redshift range. Figure \ref{fig_z_rerdm_lbg} also indicates that there is no strong dependence of $r_e/r_{\rm vir}$ in the wide luminosity range of $L_{\rm UV}\sim 0.12-10 L^*_{\rm z=3}$.

\begin{figure*}[t!]
  \begin{center}
    \includegraphics[width=120mm]{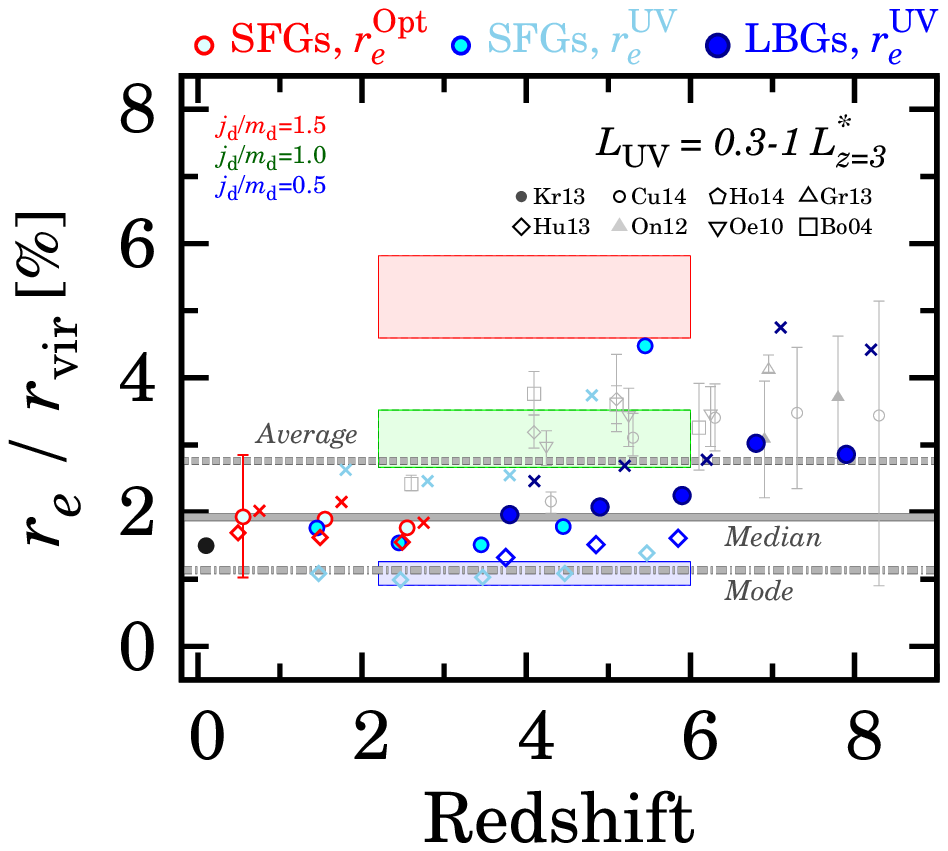}
  \end{center}
  \caption[]{{\footnotesize Comparison between our SHSR, $r_e/r_{\rm vir}$, and those of previous studies in a bin of $\sim 0.3-1\, L_{\rm UV}/L_{z=3}^*$. The symbols and lines are the same as those in Figure \ref{fig_z_re_lbg_comparison}, and we include a measurement for local galaxies with the black filled circle \citep{2013ApJ...764L..31K}. The $r_e/r_{\rm vir}$ values of the gray symbols are taken from \citet{2014arXiv1410.1535K} who compile the results of the literature. The horizontal dashed, solid, and dot-dashed lines indicate weighted means of $\left< r_e/r_{\rm vir} \right>$ of average, median, and modal values, respectively. The red, green, and blue shaded areas illustrate the regions of $j_{\rm d}/m_{\rm d}=1.5$, $1.0$, and $0.5$, respectively (see Section \ref{subsec_discuss_momentum} for details). A typical error bar in our $r_e/r_{\rm vir}$ estimates is shown at $z\sim0.5$. }}
  \label{fig_z_rerdm_comparison}
\end{figure*}

We compare our $r_e/r_{\rm vir}$ estimates with those of previous studies. Because the previous studies choose different statistics for $r_e/ r_{\rm vir}$ estimates, we present average, median, and modal $r_e/ r_{\rm vir}$ for our galaxies with $0.3-1\,L_{\rm UV}/L_{z=3}^*$ in Figure \ref{fig_z_rerdm_comparison}, together with the previous results.

For local galaxies, \citet{2013ApJ...764L..31K} obtain $r_e/ r_{\rm vir}=1.50\pm0.07$\% by the fitting of size-luminosity relations. This result of $z=0$ is consistent with our results at a similar redshift of $z\sim 0.5$ within the $1\sigma$ uncertainty (Figure \ref{fig_z_rerdm_comparison}). For high-$z$ galaxies, \citet{2014arXiv1410.1535K} calculate $r_e/ r_{\rm vir}$ values with the average statistics. In Figure \ref{fig_z_rerdm_comparison}, the gray symbols of \citealt{2014arXiv1410.1535K}'s estimates agree with blue crosses of our results. We find that the results of ours and the previous studies fall in the $r_e/ r_{\rm vir}$ range of $r_e/ r_{\rm vir}=1.0-3.5$\%, regardless of statistics choices.

Motivated by the no large evolution of $r_e/r_{\rm vir}$, we calculate $\left< r_e/ r_{\rm vir} \right>$ that is a $r_e/ r_{\rm vir}$ value weighted-averaged over $z\sim 0-8$. We obtain $\left< r_e/ r_{\rm vir} \right> =2.76\pm0.47$\%, $1.92\pm0.09$\%, and $1.13\pm0.06$\% for our $r_e/ r_{\rm vir}$ estimates of average, median, and modal statistics, respectively.

The $\left< r_e/ r_{\rm vir} \right>$ value from our average statistics results is in good agreement with that of \citet{2014arXiv1410.1535K}, $3.3\pm0.1$\%.

%%%%%%%%%%%%%%%%%%%%%%%%%%%%%%%%%%%%%%%%%%%%%%%%
\subsubsection{Dark Matter Halo and Stellar Disk}\label{subsubsec_discuss_disk}

Summarizing our observational findings for star-forming galaxies in Sections \ref{subsubsec_discuss_LND} and \ref{subsubsec_discuss_SHSR}, we identify, over cosmic time of $z\sim 0-6$, that the $r_e$ distribution is well represented by log-normal distributions, and that the standard deviation is $\sigma_{\ln{r_e}}\sim 0.45-0.75$, and that the SHSR is almost constant, $\sim 2$\%. It is interesting to compare these observational results with the theoretical predictions of the spin parameter $\lambda$ distribution of host dark halos. DM N-body simulations suggest that $\lambda$ follows a log-normal distribution with the standard deviation of $\sigma_{\ln{\lambda}}\sim0.5-0.6$ \citep[e.g., ][]{1987ApJ...319..575B,1992ApJ...399..405W,2001ApJ...555..240B}. The shape and the standard deviation of the $\lambda$ distributions are very similar to those of $r_e$. These similarities support an idea that galaxy sizes of stellar components would be related with the host DM halo kinematics. Our study has obtained this hint of $r_e$-$\lambda$ relation at the wide range of redshift, $z\sim 0-6$, that complements the previous similar claim made for $z\lesssim 3$ galaxies \citep{2014ApJ...788...28V}.

If $r_e$ values are really determined by $\lambda$ as indicated by the $r_e$ distribution properties, stellar components of the high-$z$ star-forming galaxies have dominant rotational motions that form stellar disks. In fact, according to disk formation models \citep[e.g., ][]{1983IAUS..100..391F,2002ASPC..273..289F,1980MNRAS.193..189F,1998MNRAS.295..319M}, gas receives the specific angular momentum from host DM halos through tidal interactions which make a constant SHSR similar to the one found in Section \ref{subsubsec_discuss_SHSR}. 

Moreover, in Section \ref{subsec_results_z_n}, we find that typical high-$z$ star-forming galaxies have a low S\'ersic index of $n\sim1.5$ at $z\sim 0-6$. The combination of the log-normal $r_e$ distribution, the $r_e$-$\lambda$ standard deviation similarity, and the low S\'ersic index suggests a picture that typical high-$z$ star-forming galaxies have stellar components similar to disks in stellar dynamics and morphology over cosmic time of $z\sim 0-6$.

%%%%%%%%%%%%%%%%%%%%%%%%%%%%%%%%%%%%%%%%%%%%%%%%
%%%%%%%%%%%%%%%%%%%%%%%%%%%%%%%%%%%%%%%%%%%%%%%%
\subsection{Specific Disk Angular Momentum\\ 
Inferred from the Observations and Models}\label{subsec_discuss_momentum}

As we discuss in Section \ref{sec_discussion}, a number of observational results suggest that typical high-$z$ star-forming galaxies have disk-like stellar components in dynamics and morphology at $z\sim 0-6$. Thus we compare our results with the disk formation model of \citet{1998MNRAS.295..319M}, 

\begin{equation}\label{eq_spin}
  \frac{r_e}{r_{\rm vir}} = \frac{1.678}{\sqrt{2}} \Biggl(\frac{j_{\rm d}}{m_{\rm d}}\lambda \Biggr) \frac{f_{\rm R}(\lambda, c_{\rm vir}, m_{\rm d}, j_{\rm d})}{\sqrt{f_c(c_{\rm vir})}}, 
%  \equiv \frac{1.678}{\sqrt{2}}\lambda^\prime, 
\end{equation}

\noindent where $1.678$ is a coefficient for converting the scale length of exponential disk $R_{\rm d}$ to $r_e$. The $j_{\rm d}$ ($m_{\rm d}$) value is a angular momentum (mass) ratio of a central disk to a host DM halo. The $f_c(c_{\rm vir})$ and $f_{\rm R}(\lambda, c_{\rm vir}, m_{\rm d}, j_{\rm d})$ are functions related to halo and baryon concentrations, respectively. The $c_{\rm vir}$ is the halo concentration factor. The full functional forms of $f_c(c_{\rm vir})$ and $f_{\rm R}(\lambda, c_{\rm vir}, m_{\rm d}, j_{\rm d})$ are found in \citet{1998MNRAS.295..319M}. The SHSR $r_e/r_{\rm vir}$ with a fixed $j_{\rm d}/m_{\rm d}$ shows little or no dependence on $m_{\rm d}$ and $j_{\rm d}$. If we use $\lambda$ and $c_{\rm vir}$ values well constrained by numerical simulations \citep[e.g., ][]{2002ApJ...581..799V,2009MNRAS.393.1498D,2012MNRAS.423.3018P}, we can constrain $j_{\rm d}/m_{\rm d}$.  

Figure \ref{fig_z_rerdm_comparison} presents $r_e/r_{\rm vir}$ regions corresponding to $j_{\rm d}/m_{\rm d}=0.5$, $1.0$, and $1.5$. To determine these regions, we randomly change the $\lambda$ and $c_{\rm vir}$ values within $\lambda=0.038-0.045$ \citep{2002ApJ...581..799V, 2009MNRAS.393.1498D} and $c_{\rm vir}$ ranges at $\log{M_{\rm vir}}=11-13$ M$_\odot$ in Figure 12 of \citet{2012MNRAS.423.3018P}, respectively. We also assume the conservative range of $0.05\leq m_{\rm d}\leq0.1$ \citep[e.g., ][]{1998MNRAS.295..319M}. Substituting these numbers and our results of $r_e/r_{\rm vir}$ (Section \ref{subsubsec_discuss_SHSR}) into eq. (\ref{eq_spin}), we obtain $j_{\rm d}/m_{\rm d}=0.7-0.8$. Note that our estimates of $r_e/r_{\rm vir}$ fall in $j_{\rm d}/m_{\rm d}\sim0.5-1$ at $z\sim 0-8$, regardless of the statistical choices (Figure \ref{fig_z_rerdm_comparison}).
%The uncertainties of $r_e/r_{\rm vir}$ are given by the range of $0.05\leq m_{\rm d}\leq0.1$. 

This result of $j_{\rm d}/m_{\rm d}\sim0.5-1$ indicates that a central galaxy acquire more than half of specific angular momentum from a host DM halo. Our $j_{\rm d}/m_{\rm d}$ values are comparable to the estimates with kinematical data for nearby disks \citep[$j_{\rm d}/m_{\rm d}\sim0.8$; ][]{2012ApJS..203...17R,2013ApJ...769L..26F}. Moreover, \citet{2015arXiv150301117G} predict $j_{\rm d}/m_{\rm d}\sim 1$ for $z\sim0$ late-type galaxies with the Illustris simulations \citep{2014MNRAS.444.1518V,2014Natur.509..177V,2014MNRAS.445..175G}. These independent studies for $z\sim 0$ galaxies confirm that our estimate of $j_{\rm d}/m_{\rm d}\sim0.5-1$ is correct at $z\sim 0$, and suggest that the conclusion of no significant evolution of $j_{\rm d}/m_{\rm d}$ over $z\sim 0-8$ would be reliable. \citet{2015arXiv150301117G} have revealed that galactic winds with high mass-loading factors (AGN feedback) enhance (suppress) $j_{\rm d}/m_{\rm d}$. This suggests that the no significant evolution of $j_{\rm d}/m_{\rm d}$ at $0\lesssim z\lesssim8$ would place important constraints on parameters of galaxy feedback models.

In Section \ref{subsubsec_discuss_SHSR}, we obtain that the SHSR of QGs is $\sim 0.5$\% that is about four times smaller than the one of star-forming galaxies. If we naively assume that QGs follow eq. (\ref{eq_spin}) with the one-forth of the specific angular momentum of the star-forming galaxies, we obtain $j_{\rm d}/m_{\rm d}\sim0.1-0.25$. This value is comparable to $j_{\rm d}/m_{\rm d}\sim0.1$ for nearby ellipticals in \citet{2013ApJ...769L..26F} and $j_{\rm d}/m_{\rm d}\sim0.3$ for $z\sim0$ early-type galaxies predicted in \citet{2015arXiv150301117G}. This small specific angular momentum of QGs would be explained by the loss of angular momentum via dynamical frictions during merger events and/or weak feedback \citep[e.g., ][]{2008MNRAS.389.1137S, 2008MNRAS.387..364Z}.

%% Clumpy
%%%%%%%%%%%%%%%%%%%%%%%%%%%%%%%%%%%%%%%%%%%%%%%%
\subsection{Clumpy Structures of \\
High-$z$ Star-Forming Galaxies}\label{subsec_discuss_clumpy}

Our study has shown a wide variety of morphological measurement results, supplemented by the theoretical models. It should be noted that these results are based on the structural analyses for major stellar components of the galaxies, because we mask substructures such as star-forming clumps \citep[e.g., ][]{2014arXiv1410.7398G,2014ApJ...786...15M,2014ApJ...780...77T} in our analyses. The signatures of the morphological variety could be emerged in dispersions of internal colors and SB profiles in recent structural analyses at $z\sim2-3$ \citep[e.g., ][]{2015arXiv150205713M, 2015arXiv150300722B}. Moreover, we find that the SFR SD, $\Sigma_{\rm SFR}$, increases towards high-$z$ in Figures \ref{fig_z_sfrsd_lbg} and \ref{fig_sfr_mass_sfrsd}. This fact suggests that star-forming galaxies at high-$z$ would tend to have a high gas mass density, if we assume the Kennicutt-Schmidt law \citep{1998ApJ...498..541K}. The gas-rich disks may enhance formations of star-forming clumps through the process of disk instabilities \citep[e.g., ][]{2011ApJ...733..101G}. The detailed analyses and results of clumpy stellar sub-components for our galaxy samples are presented in the paper II (T. Shibuya in preparation).

%% CONCLUSION
%%%%%%%%%%%%%%%%%%%%%%%%%%%%%%%%%%%%%%%%%%%%%%%%
%%%%%%%%%%%%%%%%%%%%%%%%%%%%%%%%%%%%%%%%%%%%%%%%
\section{SUMMARY and CONCLUSIONS}\label{sec_conclusion}

We study redshift evolution of $r_e$ and the size-relevant physical quantities such as S\'ersic index $n$, $r_e$ distribution, $r_e/r_{\rm vir}$, and $r_e-L_{\rm UV}$ relation, using the galaxy samples at $z=0-10$ made with the deep extra-galactic legacy data of $\!${\it HST}. The $\!${\it HST} samples consist of $176,152$ galaxies with a photo-$z$ at $z=0-6$ from the 3D-HST+CANDELS catalogue and $10,454$ LBGs at $z=4-10$ selected in CANDELS, HUDF09/12, and parallel fields of HFF, which are the largest samples ever used for studies of galaxy size evolution in the wide-redshift range of $z=0-10$. Our systematic size analyses with the large samples allow us to measure galaxy sizes by the same technique, and to evaluate the biases of morphological {\it K}-correction, statistics choices, and galaxy selection as well as the cosmic SB dimming. Using our galaxies at $z\sim 2-3$, we confirm that these biases are small, $\lesssim 30$\%, in the statistical sense for star-forming galaxies at high-$z$, which do not change our conclusions of size evolution.

Our findings in this study are as follows.

\begin{enumerate}
  \item The best-fit S\'ersic index shows a low value of $n\sim1.5$ for the star-forming galaxies at $z\sim0-6$. The low $n$ values indicate that a typical star-forming galaxy has a disk-like SB profile. 
  \item We derive the $r_e$-$L_{\rm UV}$ relation for star-forming galaxies over the wide-redshift range of $z=0-8$. The power-law fitting of $r_e = r_0 (L_{\rm UV}/L_0 )^\alpha$ reveals that $r_0$ values significantly decrease towards high-$z$. Similar to the evolution of $r_0$, the average, median, and modal $r_e$ values in the linear space clearly decrease from $z\sim0$ to $z\sim6$. The $r_e$ values in any statistics evolve with $r_e\propto(1+z)^{-1.0\sim-1.3}$. The slope $\alpha$ of the relation has a constant value of $\alpha=-0.27\pm0.01$ at $0\lesssim z\lesssim8$, providing an important constraint for galaxy evolution models. 
\item The SFR surface density, $\Sigma_{\rm SFR}$, increases from $z\sim0$ to $z\sim8$, while we find no stellar-mass dependence of $\Sigma_{\rm SFR}$ in this redshift range. The increase of $\Sigma_{\rm SFR}$ suggests that high-$z$ star-forming galaxies would have a gas mass density higher than low-$z$ star-forming galaxies on average, if one assumes that the Kennicutt-Schmidt law does not evolve significantly by redshift. 
  \item We identify a clear positive correlation between $r_e$ and $\beta$ for star-forming galaxies at $z\sim0-7$ in the luminosity range of $0.3-1.0\,L^*_{\rm z=3}$. This is explained by a simple picture that galaxies with young stellar ages and/or low metal+dust contents typically have a small size. 
  \item The $r_e$ distribution of UV-bright star-forming galaxies is well represented by log-normal functions. The standard deviation of the log-normal $r_e$ distribution, $\sigma_{\ln{r_e}}$ is $\sim0.45-0.75$, and $\sigma_{\ln{r_e}}$ does not significantly change at $z\sim0-6$. Note that the structure formation models predict that the distribution of a DM spin parameter $\lambda$ follows a log-normal distribution with the $\lambda$-distribution's standard deviation of $\sigma_{\ln{\lambda}}\sim 0.5-0.6$. The distribution shapes and standard deviations of $r_e$ and $\lambda$ are similar, supporting an idea that galaxy sizes of stellar components would be related with the host DM halo kinematics. 
  \item Combining our stellar $r_e$ measurements with host DM halo radii, $r_{\rm vir}$, estimated from the abundance matching study of Behroozi et al., we obtain a nearly constant value of $r_e/r_{\rm vir}=1.0-3.5$\% at $0\lesssim z\lesssim 6$ in any statistical choices of average, median, and mode. 
  \item The combination of the log-normal $r_e$ distribution with $\sigma_{\ln{r_e}} \sim0.45-0.75$, and the low S\'ersic index suggests a picture that typical high-$z$ star-forming galaxies have stellar components similar to disks in stellar dynamics and morphology over cosmic time of $z\sim 0-6$. If we assume the disk formation model of \citet{1998MNRAS.295..319M}, our $r_e/r_{\rm vir}$ estimates indicate that a central galaxy acquires more than a half of specific angular momentum from their host DM halo,  $j_{\rm d}/m_{\rm d}\simeq 0.5-1$. 
 \end{enumerate}

These results are based on galaxies' major stellar components, because we mask galaxy sub-structures such as star-forming clumps in our analyses. The detailed analyses and results for the clumpy stellar sub-components are presented in the paper II (T. Shibuya in preparation). We expect that future facilities such as the {\it James-Webb Space Telescope}, {\it Wide-Field Infrared Survey Telescope},  {\it Wide-field Imaging Surveyor for High-redshift} telescope, and 30-meter telescopes will obtain deep NIR images with a high spatial resolution, PSF FHWM of $\lesssim 0.\!\!^{\prime\prime}1-0.\!\!^{\prime\prime}2$, for a large number of galaxies at $z\sim10$ and beyond. Surveys with these facilities will reveal when the size-luminosity relation emerges  and whether first galaxies fall in the extrapolation of the $r_e$ evolution and the nearly constant relation of $r_e/r_{\rm vir}=1.0-3.5$\% at $z\sim 0-8$ and $0-6$, respectively, that we find in this study.

%% ACKNOWLEDGMENTS
%%%%%%%%%%%%%%%%%%%%%%%%%%%%%%%%%%%%%%%%%%%%%%%%
%%%%%%%%%%%%%%%%%%%%%%%%%%%%%%%%%%%%%%%%%%%%%%%%

\acknowledgments

We thank the anonymous referee for constructive comments and suggestions. We would like to thank Anahita Alavi, Steven Bamford, Rychard J. Bouwens, Marcella C. Carollo, Emma Curtis-Lake, Michael Fall, Ryota Kawamata, Chervin Laporte, Z. Cemile Marsan, Andrew Newman, Carlo Nipoti, Tomoki Saito, Kazuhiro Shimasaku, Genel Shy, Ignacio Trujillo, Masayuki Umemura, Arjen van der Wel, and Suraphong Yuma for their encouragement and useful discussion and comments. We thank Yoshiaki Ono for kindly providing us a part of HUDF09-P1 and P2 images. This work is based on observations taken by the 3D-HST Treasury Program (GO 12177 and 12328) and CANDELS Multi-Cycle Treasury Program with the NASA/ESA HST, which is operated by the Association of Universities for Research in Astronomy, Inc., under NASA contract NAS5-26555. Support for this work was provided by NASA through an award issued by JPL/Caltech. This work was supported by World Premier International Research Center Initiative (WPI Initiative), MEXT, Japan, KAKENHI (23244025) and (21244013) Grant-in-Aid for Scientific Research (A) through Japan Society for the Promotion of Science (JSPS), and an Advanced Leading Graduate Course for Photon Science grant.

{\it Facilities:} \facility{HST (ACS, WFC3)}.

%% REFERENCE
%%%%%%%%%%%%%%%%%%%%%%%%%%%%%%%%%%%%%%%%%%%%%%%%
%%%%%%%%%%%%%%%%%%%%%%%%%%%%%%%%%%%%%%%%%%%%%%%%

\bibliographystyle{apj}
\bibliography{reference}

\clearpage

\end{document}